\documentclass[reprint, amsmath,amssymb, aps,pra]{revtex4-1}
\usepackage{stackengine}
\usepackage{booktabs}
\usepackage{float}

\usepackage[caption=false]{subfig}
\usepackage{graphicx}
\usepackage{bm}
\usepackage{hyperref}
\usepackage{natbib}
\usepackage[usenames,dvipsnames]{color}
\usepackage[usenames,dvipsnames,svgnames,table]{xcolor}
\usepackage{physics}

\usepackage{comment}
\bibpunct{\color{blue}[}{\color{blue}]}{,}{n}{}{;}
\hypersetup{
  colorlinks,
  citecolor=blue,
  linkcolor=blue,
  urlcolor=blue
  }

\newlength{\mysize}

\begin{document}
\preprint{APS/123-QED}
\title{ Local and non-local quantum transport  due to Andreev bound states \\ in finite Rashba nanowires with superconducting and normal sections}
\author{Richard Hess}
\author{Henry F. Legg}
\author{Daniel Loss}
\author{Jelena Klinovaja}%
\affiliation{%
 Department of Physics, University of Basel, Klingelbergstrasse 82, CH-4056 Basel, Switzerland }%
 \date{\today}
 \begin{abstract}
We analyze Andreev bound states (ABSs) that form in normal sections of  a Rashba nanowire that is only partially covered by a superconducting layer. These ABSs are localized close to the ends of the superconducting section and can be pinned to zero energy over a wide range of magnetic field strengths  even if the nanowire is in the non-topological regime. For  finite-size nanowires (typically $\lesssim 1$ $\mu$m in current experiments), the ABS localization length is comparable to the length of the nanowire.  The  probability density of an ABS is therefore non-zero throughout the nanowire and differential-conductance calculations reveal a correlated zero-bias peak (ZBP) at both ends of the nanowire. When a second normal section  hosts an additional ABS at the opposite end of the superconducting section, the combination of the two ABSs can mimic the closing and reopening of the bulk gap in local and non-local   conductances accompanied by the appearance of the ZBP. These signatures are reminiscent of those expected for Majorana bound states (MBSs) but occur here in the non-topological regime. Our results demonstrate that  conductance measurements of correlated ZBPs at the ends of a typical superconducting nanowire or an apparent closing and reopening of the bulk gap in the local and non-local  conductance are not conclusive indicators for the presence of MBSs.
\end{abstract}

\maketitle

\section{Introduction}
Majorana bound states (MBSs) have been of significant interest in condensed matter physics for over two decades, largely due to their potential application as topological qubits  \cite{kitaev2001unpaired,Fujimoto2008Topological,Sato2009NonAbelian,
Volovik2009Fermion,OregHelical2010,LutchynMajorana2010,
Sato2010NonAbelian}. 
The prospective utilization of MBSs in quantum computation stems from their non-Abelian braiding statistics  \cite{Moore1991Nonabelions,Volovik1999Fermion,Read2000Paired,
Senthil2000Quasiparticle,Ivanov2001NonAbelian,Nayak2008Non,
alicea2012new,beenakker2013search}. 
Despite this intense interest there has been no conclusive experimental observation of these exotic properties to date.

\begin{figure}[t]
\subfloat{\label{figNanoWireScatteringRegionLeftNormalSectionBarrier}\stackinset{l}{-0.03in}{t}{-0.in}{(a)}{\includegraphics[width=1\columnwidth]{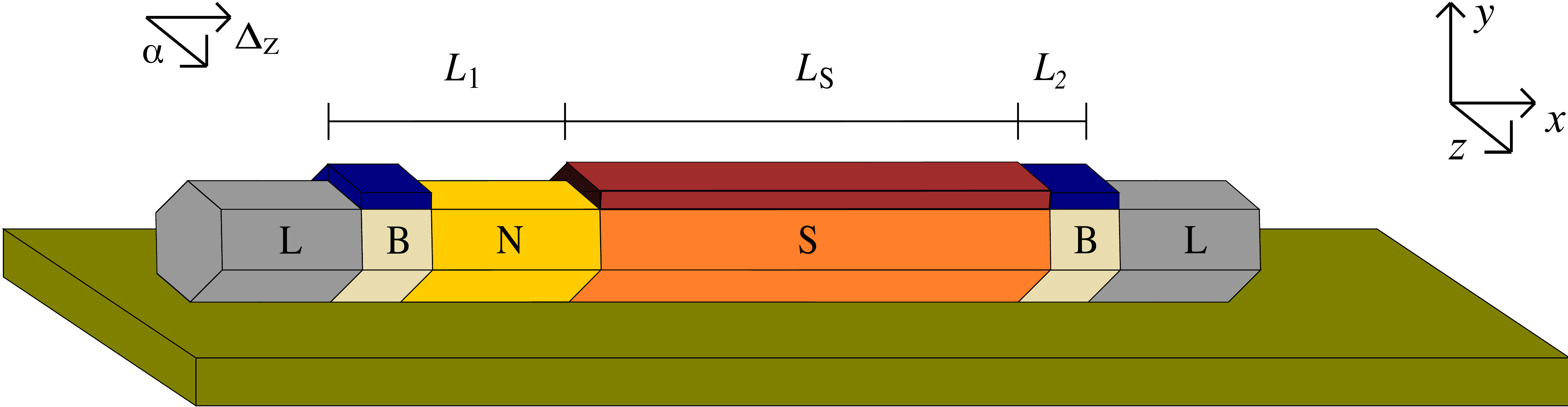}} 
}
\\
\subfloat{\label{figNanoWire1}\stackinset{l}{-0.03in}{t}{0.in}{(b)}{\includegraphics[width=1\columnwidth]{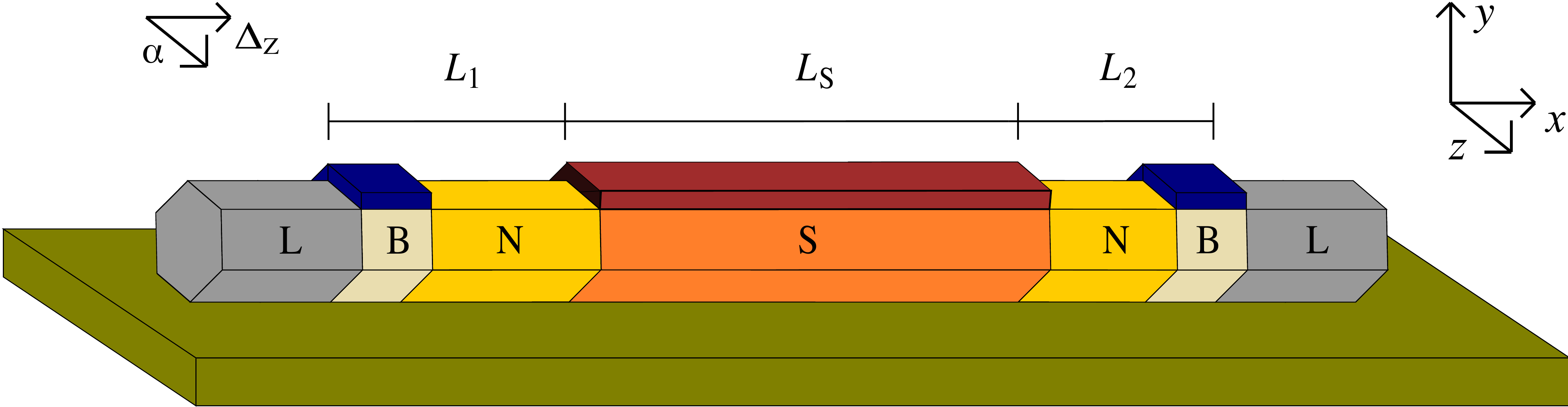}}
}
 \caption{Different configurations of a nanowire setup considered in this work: The semiconducting  nanowire is aligned along the $x$-axis. The Rashba vector points in $z$-direction and the applied magnetic field in $x$-direction. A grounded $s$-wave superconductor (dark red) covers a section of length $L_S$ (orange),  locally  inducing superconductivity  via the proximity effect. (a) Only a left section (yellow) or (b) both left and right sections on both ends of length $ L_1 $ and $ L_2 $ are uncovered  by the superconductor and remain normal.  
 Leads  (gray) are attached on the left and right end to  measure the differential conductance of the system. Tunnel barriers (beige) at the ends of the normal sections  can be used to control differential conductance, the height of these tunnel barriers is tuned by local contacts (dark blue).}
 \label{figConfigurationsNanowire}
 \end{figure}

The most mature experimental platform expected to host MBSs are Rashba nanowires (see Fig.~\ref{figConfigurationsNanowire}), where the key differential-conductance signature associated with MBSs is a zero-bias peak (ZBP) that is stable for a wide range of magnetic field strengths. A ZBP is, however, by itself not a unique fingerprint of MBSs. Previously it was suggested that additional local  conductance features can clarify the origin of such a ZBP, namely the quantization of the peak height at $2e^2/h$ \cite{law2009majorana, akhmerov2009electrically,Flensberg2010Tunneling,Wimmer2011Quantum,Chevallier2016Tomography} and oscillations around zero energy that originate from the overlap of the two MBS wavefunctions at either end of the nanowire \cite{Prada2012Transport,sarma2012splitting,Rainis2013Towards, Dmytruk2018Suppresion, Fleckenstein2018Decaying}.  The ZBPs and their oscillations have been observed in past experiments \cite{Sasaki2011Topological,MourikSignatures2012,Deng2012Anomalous,Das2012Zero,Churchill2013Superconductor,AlbrechtExponential2016,schneider2021controlled}, while quantization of the ZBP has not been observed. Recently it has been suggested \cite{Rosdahl2018Andreev,pikulin2021protocol} that the next generation of Rashba nanowire systems, three-terminal devices, could elucidate whether a given ZBP stems from the presence of MBSs by observing additional auxiliary features in the local and non-local  differential  conductances. For example, such devices could observe correlations between ZBPs at both ends of the nanowire and the 
closing and reopening of the bulk-gap that should accompany the transition to topological superconductivity.

Additional signatures of MBSs beyond a simple ZBP are necessary because topologically trivial states such as Andreev bound states (ABSs) \cite{DeGennes1963Elementary,Andreev1964Thermal,Caroli1964Bound,Andreev1966Electron, Shiba1968Classical, Rusinov1969On,Sauls2018Andreev,Prada2020From} can generate conductance features similar to those expected from MBSs and therefore strongly challenge the interpretation of experimental observations \cite{kells2012Near,Lee2012Zero,Cayao2015SNS,Ptok2017Controlling,Liu2017Andreev,
reeg2018zero,Penaranda2018Quantifying,moore2018two,Vuik2019Reproducing,Woods2019Zero,
Liu2019Conductance,Chen2019Ubiquitous,Alspaugh2020Volkov,Prada2020From,
Juenger2020Magnetic,valentini2020nontopological}. For instance, it has been shown that the energy of an ABS in a non-topological system can be pinned close to zero over a wide range of magnetic field strengths when a resonance condition for the strength of the spin-orbit interaction (SOI) is fulfilled \cite{reeg2018zero}.  In transport experiments, this resonance is broadened by finite temperature and the coupling to external leads.
Such ABSs can therefore produce ZBP features in the  conductance even in systems that are topologically trivial at all magnetic field strengths. The pinning of trivial ABSs close to zero energy can also originate from smooth parameter profiles of the chemical potential and the superconducting gap \cite{moore2018two,Penaranda2018Quantifying,Vuik2019Reproducing}, such that  a short section of the nanowire is nominally in the topological regime. Such zero-energy states, observed in the trivial phase of the bulk of the nanowire, are known as quasi-Majorana bound states (quasi-MBSs) and their zero-bias pinning is in fact also stable against changes of SOI strength or tunnel barrier gate voltage.

Previous devices focussed on local measurements on a single end of a nanowire. Such measurements can already provide additional indicators that could clarify the origin of a ZBP. One example is the oscillations around zero energy expected due to the hybridization of MBSs at either end of a finite nanowire \cite{Prada2012Transport,sarma2012splitting,Rainis2013Towards, Dmytruk2018Suppresion}. Such oscillations should have an increasing amplitude when magnetic field strength is increased or nanowire length decreased.  In contrast to this expectation, several experiments observed oscillations with an amplitude which decays as the magnetic field is increased \cite{AlbrechtExponential2016,Farrell2018Hybridization,Shen2018parity}. Although there are proposed explanations for this behaviour such as orbital effects \cite{Dmytruk2018Suppresion} or a step-like profile of the Rashba SOI strength \cite{cao2019decays}, even in such scenarios the parameter window for a decay in the amplitude of oscillations is rather small and therefore the experimentally observed behaviour is likely the result of trivial states. In addition, recent theoretical works have shown that even the quantization of a ZBP at one end of the nanowire is not an exclusive property of MBSs \cite{Vuik2019Reproducing, pan2020generic}. As such, while conductance oscillations and even quantization can provide limited additional evidence for the potential presence of MBSs, they are not sufficient for an unambiguous identification of topologically protected states.
 
Given the ambiguous origins of previous experimental observations from the single end of a nanowire, in the absence of braiding experiments, further signatures in conductance are necessary to improve the classification of ZBPs in the next generation of Rashba nanowire systems. For instance, this can be achieved by considering non-local correlation properties of  MBSs in three-terminal devices \cite{Entin2008Conductance,LobosTunneling2014,Gramich2017Andreev,
Rosdahl2018Andreev,Zhang2019Next,DanonNonlocal2020,Melo2021Conductance, Pan2021Three,pikulin2021protocol}. MBSs should be localized at the opposite ends of a superconducting Rashba nanowire and therefore conductance measurements on both ends should reveal ZBPs. Furthermore, three-terminal experiments enable the measurement of non-local  conductances
which can  indicate the  bulk-gap closing and reopening and therefore go beyond local properties. Recently, it was highlighted in Ref.~\cite{Rosdahl2018Andreev}  that the exponential decay of sub-gap states into the bulk of the nanowire makes the non-local  conductance an ideal tool for distinguishing between trivial and topological phases in nanowires which are much longer than the localization length of such sub-gap states. Recent experiments have been performed on three-terminal devices  \cite{Menard2020Conductance,puglia2020closing,yu2020non} but so far did not find clear signatures of MBSs. 

In this paper we focus mainly on  {\it non-topological} three-terminal  junctions consisting of a partially proximitized Rashba nanowire where the normal sections  can host an ABS. 
We consider normal-superconducting (NS)  and normal-superconducting-normal (NSN) junction setups.
In contrast to previous works, we examine the case where the  ratio between the length of the superconducting section and the localization length of ABSs is small. This regime is of present experimental relevance and the nanowire lengths as well as the superconducting gaps we consider will be comparable to current setups where typical lengths are between $0.4$ $\mu$m   \cite{yu2020non} and $1$ $\mu$m  \cite{puglia2020closing}. The nanowire length is limited by the requirement of working in the ballistic regime to avoid disorder effects, which were shown to be harmful for the observation of topological phases. In the short-nanowire regime the wavefunction of a trivial ABS  leaks from one end of the nanowire to the opposite end. When the parameters of the ABS are close to the resonance condition from Ref.~\cite{reeg2018zero}, the  energy of the ABS is pinned close to zero over a wide range of magnetic field strengths. Our calculation of the differential conductance confirms that in such a scenario correlated ZBPs of a trivial origin appear at both ends of the nanowire. We find that the same effect can occur for quasi-MBSs in topological nanowires. 

We also examine  the consequences of the presence of a second normal section hosting an additional ABS on the other side of the superconducting section. Such NSN junctions with two normal sections are expected to naturally occur in three-terminal devices available experimentally.  We find that the appearance of the second ABS can further complicate the interpretation of experimental signatures. Not only is the second ABS also visible in the non-local  conductance but the combination of the two ABSs at either end of the nanowire can generate a conductance feature that is reminiscent of the bulk-gap edge undergoing a closing and reopening process that should accompany a topological phase transition. 

Our findings show that, while three-terminal devices can potentially provide additional insights into the origins of ZBPs, correlated zero-bias peaks at both ends of superconducting sections of Rashba nanowires and the apparent observation of the closing and reopening of the bulk band gap with increasing magnetic field strength do not suffice as unambiguous additional indicators for the presence of MBSs in nanowires of the lengths used in current experimental devices.

The paper is organized as follows. In Sec.~\ref{Sec:ModelNanowires} we define the model to describe a non-topological and a topological nanowire containing trivial zero-energy ABSs or quasi-MBSs, respectively.  In Sec.~\ref{SecNanowireSharpProfiles}, we discuss features in  the differential conductance  arising due to the presence of a single ABS hosted in the, say, left normal section of a non-topological nanowire. 
Here we show that as the ratio between the length of the superconducting section and the localization length of the ABS is decreased, the probability density of the ABS on the right side of the nanowire increases and, as a result, the ABS also becomes visible in the local  conductance measured at the right end of the nanowire.
Moreover, we examine the case of an NSN junction with two normal sections, one at each end of the non-topological proximitized nanowire, and show that this setup can mimic the signatures of a topological phase transition in transport measurements, despite the trivial nature of the ABSs.  Section ~\ref{sec:NanoWireSmooth} focuses on the topological nanowire and addresses features arising due to the presence of quasi-MBSs in the left and right local  conductance. It is shown again that if the ratio between the superconducting section and the localization length of the quasi-MBS is small, then correlated zero-bias  peaks appear at both ends.
Furthermore, we examine the non-local  differential conductance via the  bulk states undergoing the bulk-gap closing and reopening process when two normal sections at each end of the topological nanowire both host quasi-MBSs.
Finally, we discuss the impact of our results on the interpretation of present-day three-terminal experiments in Sec.~\ref{Sec:Conclusion}.  In  App.~\ref{App:DiffCond} we describe numerical approaches used to model transport experiments. We compare the conductance pattern of the non-topological nanowire with  the conductance pattern of a uniform topological nanowire in App.~\ref{App:TopoNanowireAdditionalMaterial}. The effect of  strong broadening of finite-energy peaks  is discussed in App.~\ref{App:Broadening}.  In App. \ref{App:WavefunctionNoGapClosing} we study the  bulk  wavefunctions of a topological nanowire with quasi-MBSs on both ends. Finally, App. \ref{App:QuasiMBSAndABS} deals with the conductance pattern of a nanowire  hosting quasi-MBSs on the left end and an ABS on the right end.

\section{Model of the nanowire \label{Sec:ModelNanowires}}
We consider a one-dimensional (1D) semiconducting nanowire aligned along the $x$-direction.  The system is subjected to a magnetic field, which is applied parallel to the nanowire axis. This magnetic field results in a Zeeman energy  of strength $\Delta_Z(x)$.  The nanowire is partially covered by an $s$-wave superconductor, resulting in a proximity-induced superconducting gap $\Delta(x)$ in a section of length $L_S$.  This grounded superconducting section is centered between a left  and a right normal section of length $L_1$ and $L_2$, respectively.  The Rashba SOI of  strength $\alpha(x)$ is position-dependent and the corresponding SOI vector points in $z$-direction. The effective 1D lattice Hamiltonian is given by 
\begin{align}
H=&\sum_{n=1}^{N} \Bigg[\sum_{\sigma,\sigma'}c_{n,\sigma}^{\dagger}( \lbrace t_{n+\frac{1}{2}}+t_{n-\frac{1}{2}}-\mu_n+\gamma_n\rbrace\delta_{\sigma \sigma'}\nonumber \\
&+\Delta_{Z,n}\sigma_{\sigma \sigma'}^x) c_{n,\sigma'}\nonumber \\
&-\Bigg(\sum_{\sigma,\sigma'}c_{n,\sigma}^{\dagger}\left\lbrace t_{n+\frac{1}{2}}\delta_{\sigma \sigma'}-i\alpha_{n+\frac{1}{2}}\sigma_{\sigma \sigma'}^z\right\rbrace c_{n+1,\sigma'}\nonumber \\ &+\Delta_n c_{n,\downarrow}^{\dagger}c_{n,\uparrow}^{\dagger}  +H.c. \Bigg) \big. \Bigg] \label{eq:effHam},
\end{align}
where $c_{n,\sigma}^{\dagger}$  and  $c_{n,\sigma}$ creates and annihilates  an electron of  spin $\sigma=\uparrow,\downarrow$ at the lattice site $n$, respectively.  The total number of sites is given by $N=N_1+N_S+N_2$, where $N_1=L_1/ a $, $N_2=L_2/a$, and $N_S=L_S/ a$, where $a$ is an effective lattice constant. In addition,  $t_n$ and $\mu_n$ denote the nearest neighbor tunneling matrix element and the chemical potential, respectively. Furthermore, $ \delta_{\sigma \sigma'}  $ denotes the Kronecker delta. Normal leads are attached on the left and right ends of the nanowire. The leads are modeled by the same Hamiltonian as the normal sections. The chemical potentials  $\mu_L$ and $\mu_R$ of the left and right normal lead  are adjusted to account for  a possible difference between lead and nanowire. Additionally, we introduce tunnel barriers between  the leads and  the nanowire, these barriers of length $L_{B,1}$ and $L_{B,2}$ are constituent parts of the normal sections of length $L_1$ and $L_2$, see Fig.~\ref{figNanoWire1}. The height $\gamma_n$  of the tunnel barrier at site $n$ is used to control the coupling between the system and the leads, it therefore controls the conductance value. We will focus on two setups, which we refer to as the `non-topological' and `topological' nanowire. Both of these systems can host ABSs that are pinned to zero energy, however, the mechanism fixing the ABS energy to zero is different for the two nanowire types. These specific parameter configurations for the non-topological and  topological cases are described in the following subsections \ref{SubSec:NonTopSys} and \ref{SubSec:TopSys}. \\

\begin{figure}[t]
\subfloat{\label{figParameterProfileSharpOneQDot}\stackinset{l}{-0.03in}{t}{-0.03in}{(a)}{\includegraphics[width=0.48\columnwidth]{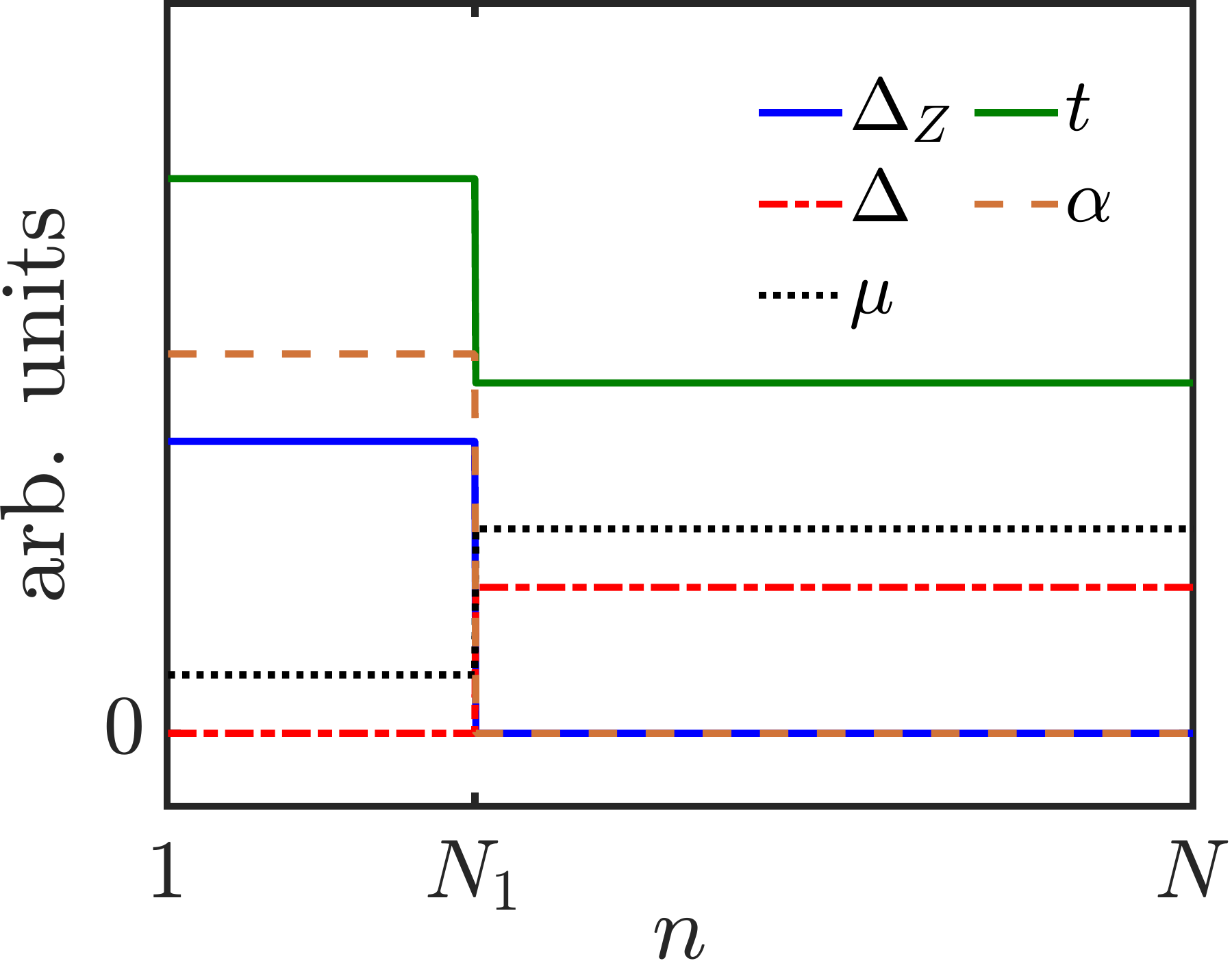}}
}
\hfill
\subfloat{\label{figParameterProfileSharpTwoQDot}\stackinset{l}{-0.03in}{t}{-0.03in}{(b)}{\includegraphics[width=0.48\columnwidth]{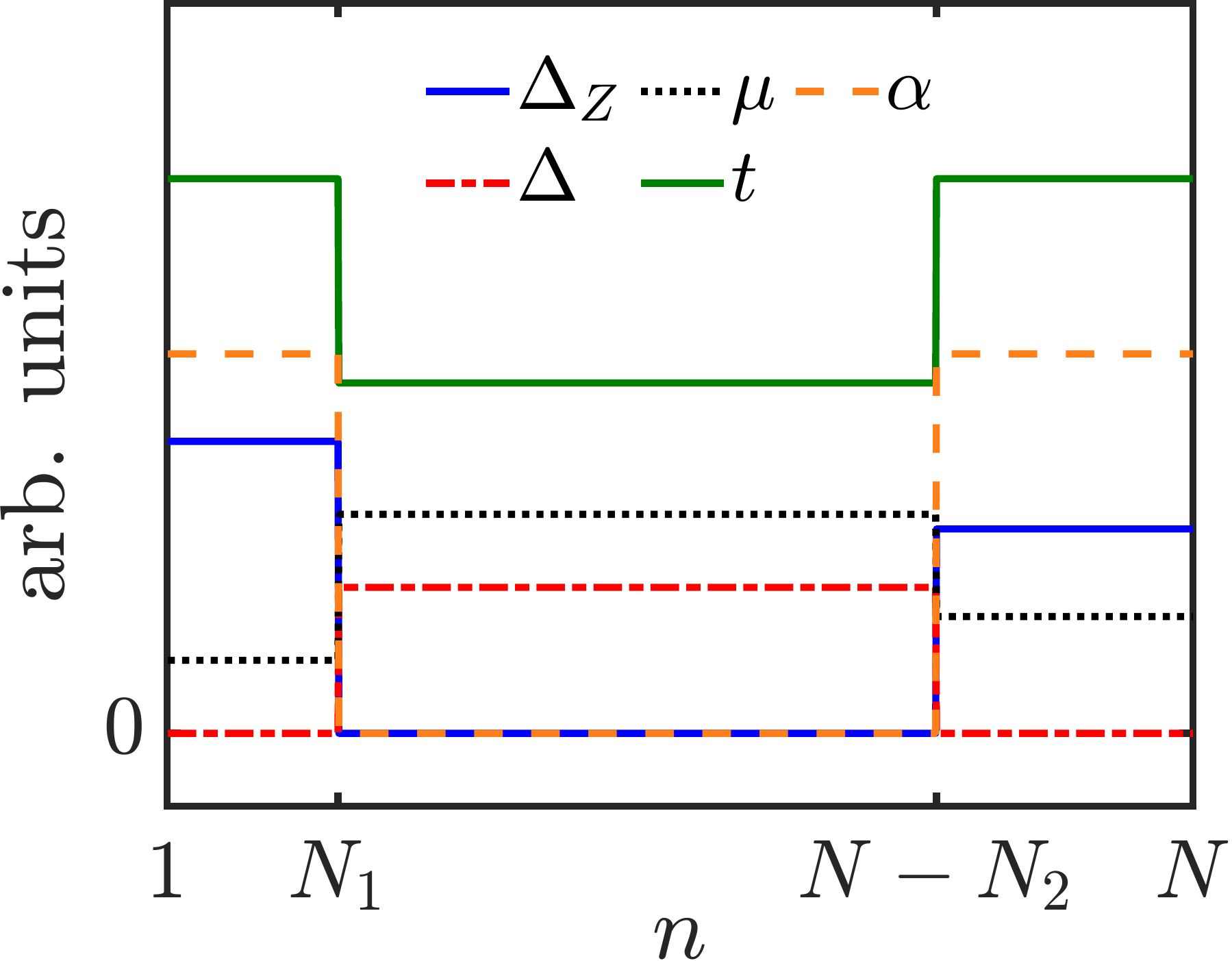}}
}
\vspace{-0.1in}
\subfloat{\label{figParameterProfileSmoothOneQDot}\stackinset{l}{-0.03in}{t}{-0.03in}{(c)}{\includegraphics[width=0.48\columnwidth]{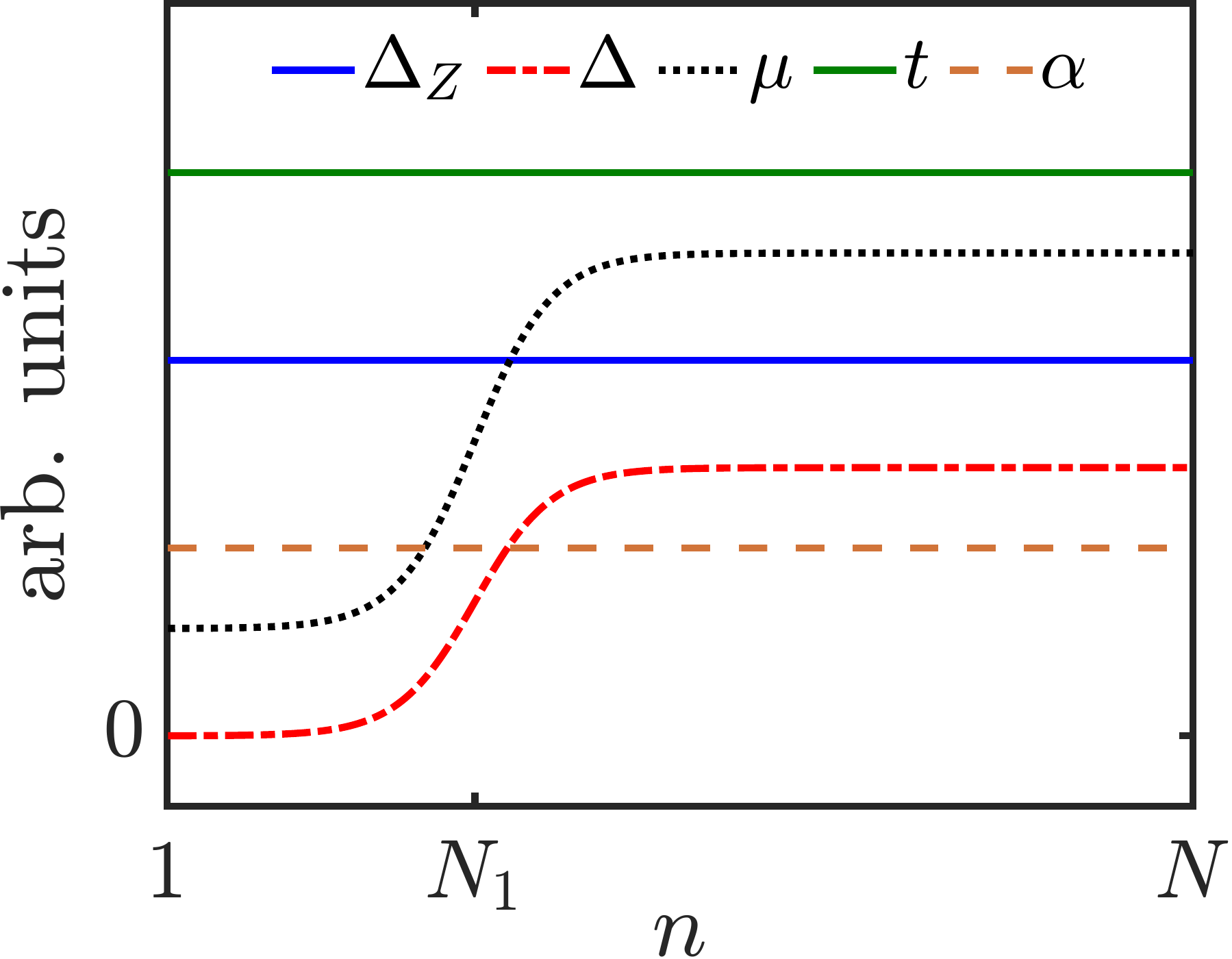}}
}
\hfill
\subfloat{\label{figParameterProfileSmoothTwoQDot}\stackinset{l}{-0.03in}{t}{-0.03in}{(d)}{\includegraphics[width=0.48\columnwidth]{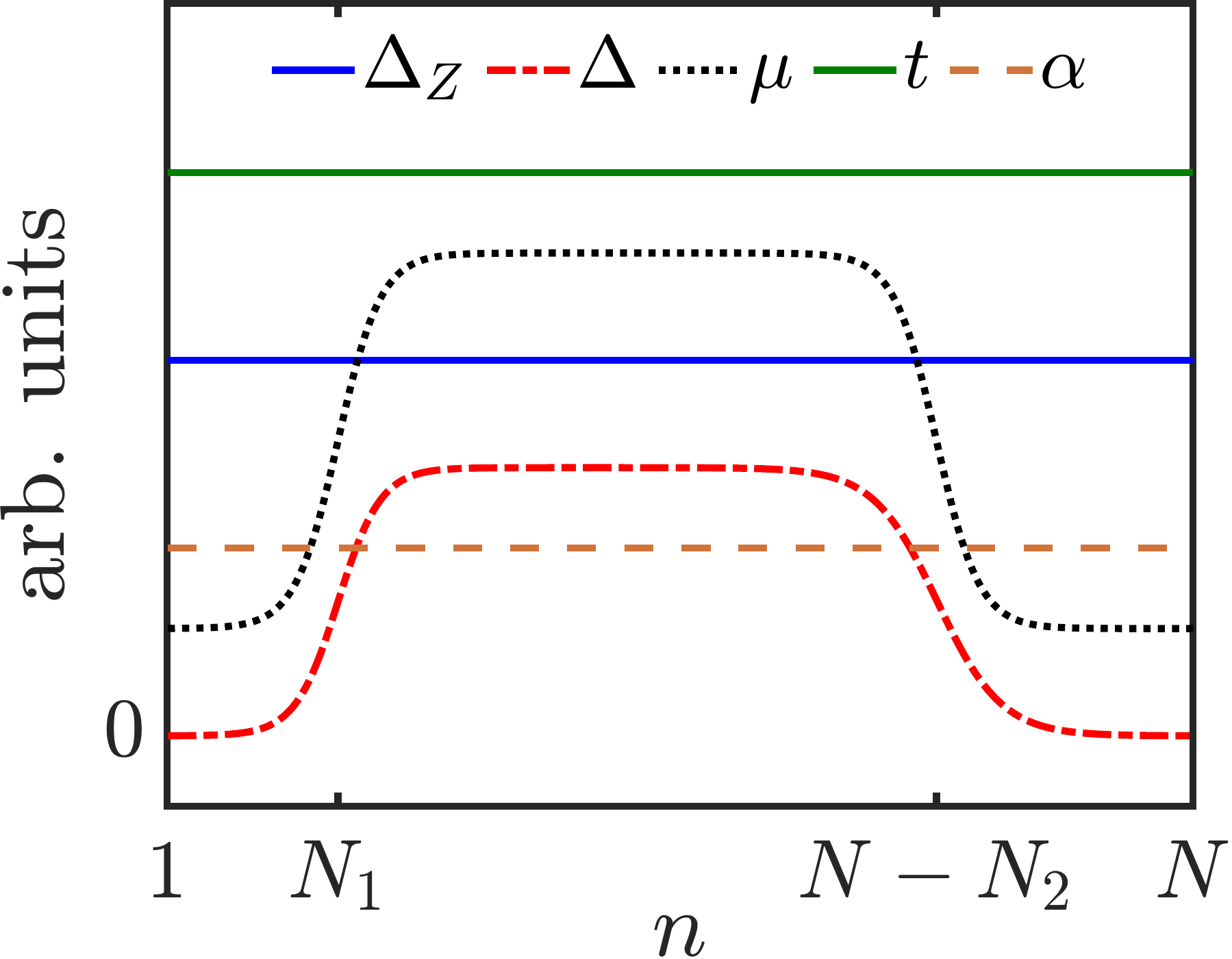}}
}
 \caption{Parameter profiles in the  nanowire: superconducting gap $\Delta $ (dashed-dotted red), chemical potential $ \mu $ (dotted black),  Zeeman energy $\Delta_Z$ (solid blue), SOI strength $\alpha$ (dashed orange), and tunneling matrix element $t$ (green) in arbitrary units.  Non-topological nanowire (first row):  Both the SOI and the Zeeman energy are suppressed in the superconducting section.  When a resonance condition in the normal part is satisfied,  the lowest ABS is pinned to zero energy.  Topological nanowire (second row): Smooth parameter profiles lead to a zero-energy pinning of the lowest ABS in the topologically trivial bulk-phase. 
 Left column: NS-junction. Right Column: NSN-junction.}
 \label{figParameterProfileSharp}
 \end{figure}

\subsection{Non-topological nanowire} \label{SubSec:NonTopSys}

In this section we specify the profiles of the parameters that enter the Hamiltonian, $H $, given in Eq.~\eqref{eq:effHam} for the non-topological nanowire.  We define the boundary between the left normal section ($\text{N}_1$)  and the superconducting section (S)  as  $N_b=N_1+1/2$ and similarly the boundary between S and the right normal section ($\text{N}_2$)  as  $N_b'=N_1+N_S+1/2$. 
The non-uniform system parameters entering the Hamiltonian $H$ have the following structures: The tunneling matrix element is defined as
\begin{align}
t_{n}= & t_{1}\Theta(N_b-n)+t_S\left[\Theta(n-N_b)-\Theta(n-N_b')\right]\nonumber \\&+t_{2}\Theta(n-N_b')
\end{align} 
and is constructed out of the tunneling matrix elements $t_{1}=t_2$ in $\text{N}_1$  and $\text{N}_2$ and the tunneling matrix element  $t_S$ in S. We define the Heaviside function $ \Theta(n) $ with $\Theta(0)=1/2$ throughout.   The difference between the tunneling matrix elements of the superconducting and the normal sections arises due to the mass renormalization inside the superconducting section caused by metallization effects induced by the thin superconducting shell \cite{Reeg2017Finite,Reeg2018Metallization,Reeg2018Proximity,Woods2019Electronic,
Winkler2019Unified,Kiendl2019Proximity}.  
The chemical potential has a similar structure 
\begin{align}
 \mu_n=&\mu_{1}\Theta(N_b-n)+\mu_s\left[\Theta(n-N_b)-\Theta(n-N_b')\right]\nonumber \\&+\mu_{2}\Theta(n-N_b'),
\end{align}
where $\mu_{1}$ and $\mu_{2}$ denote the chemical potential in  the normal sections and $ \mu_S $ the chemical potential in the superconducting section. Since the magnetic field suppresses the bulk-gap of the parent  superconductor,  the superconducting gap decreases with increasing Zeeman energy and vanishes at the critical field strength $\Delta_Z^c$:
\begin{align}
\Delta=\Delta_0\sqrt{1-(\Delta_Z/{\Delta_Z^c)^2}}, \label{de}
\end{align}
where the maximal value is given by $ \Delta_0$. Therefore, the superconducting gap has the following profile  
\begin{align}
\Delta_n=\Delta \left[\Theta(n-N_b)-\Theta(n-N_b')\right],
\end{align}
The superconducting  gap is zero in $\text{N}_1$ and $\text{N}_2$. In contrast,  the Zeeman energy and Rashba SOI are non-zero only in the normal sections and are defined as
\begin{align}
&\Delta_{Z,n}=\Delta_{Z}\Theta(N_b-n)+\Delta_{Z}\Theta(n-N_b'),\\
&\alpha_n=\alpha_{1}\Theta(N_b-n)+\alpha_{2}\Theta(n-N_b').
\end{align} 
Here, the SOI strengths $\alpha_{1}$ and  $\alpha_{2}$ could be different \cite{Klinovaja2015Fermionic}. The SOI energy  is given by $E_{\text{so},i}=\alpha_i^2/t_i$.  In Fig.~\ref{figParameterProfileSharp} we show examples of the profiles of the superconducting gap, the Zeeman energy, the chemical potential, the tunneling matrix element, and the Rashba SOI strength for an NS and an NSN junction. 
Tunnel barriers are described by 
\begin{align}
 \gamma_n=&\gamma_{1}\Theta(N_{B,b}-n)+\gamma_{2}\Theta(n-N_{B,b}'),\label{eq:barrier}
\end{align} 
where $\gamma_1$ and $\gamma_2$ denote the height of the left and right tunnel barriers and $N_{B,b}=N_{B,1}+1/2 $ and $N_{B,b}'=N-N_{B,2}+1/2$ are the positions at which the left tunnel barrier ends and the right tunnel barrier starts, respectively. Here we defined $N_{B,i}$ via $L_{B,i}=N_{B,i}a$.  
We note that  the topological phase cannot be achieved in this setup because the Zeeman energy and the Rashba SOI vanish in the superconducting section. We therefore refer to this system as {\it non-topological  nanowire}.

\subsection{Topological  nanowire}\label{SubSec:TopSys}
The second system under consideration is a nanowire in which the chemical potential and the superconducting gap change smoothly.  These smooth parameter variations can generate an ABS which, as in the non-topological nanowire, sticks to zero energy over a wide range of Zeeman energies in the trivial regime inside the superconducting section \cite{kells2012Near,moore2018two,Penaranda2018Quantifying,Vuik2019Reproducing}.  In this case, nominally, the system enters the topological phase locally at the short segment between the normal and superconducting sections. However, the length of this segment is much shorter than the localization length of potential MBSs, such that only quasi-MBSs can appear in the spectrum if certain conditions are satisfied. The spatial dependence of parameters is modelled by the function 
\begin{align}
 \Omega_{\lambda}(n,N_i)=1/2[1+\tanh(\lbrace n-N_i \rbrace/\lambda)] ,
\end{align} where $\lambda$ parametrizes the smoothness (see Fig.~\ref{figParameterProfileSharp}c-d).  The exact form of the function is not relevant for the appearance of  quasi-MBSs rather it is the smoothness itself that determines the presence of quasi-MBSs. The  superconducting gap (chemical potential) profile  is characterized by the parameter $\lambda_{S,L/R}$ ($ \lambda_{L/R} $), which can take different values on the left and the right side of the nanowire:
\begin{align}
&\Delta_n=\Delta_0\left[\Omega_{\lambda_{S,L}}(n,N_1)-\Omega_{\lambda_{S,R}}(n,N_1+N_S+1)\right],\\
&\mu_n= \mu_1+(\mu_S-\mu_1)\Omega_{\lambda_{L}}(n,N_1)\nonumber \\
&\hspace{50pt}+ (\mu_2-\mu_S)\Omega_{\lambda_{R}}(n,N_1+N_S+1),
\end{align}
In contrast to the previous section, here, we use a superconducting gap that is independent of the Zeeman energy. For the case of a single normal section on the left and a tunnel barrier only on the right, we choose the  profiles:
 \begin{align}
&\Delta_n=\Delta_0\Omega_{\lambda_{S,L}}(n,N_1) \Theta(N_{B,b}'-n),\\
&\mu_n= \left[\mu_1+(\mu_S-\mu_1)\Omega_{\lambda_{L}}(n,N_1)\right]\Theta(N_{B,b}'-n)\nonumber \\
&+\mu_2\Theta(n-N_{B,b}').
\end{align} 
The tunnel barriers are modeled in the same manner as in the non-topological system, see Eq.~\eqref{eq:barrier}.  The remaining parameters  are chosen to be uniform:
 \begin{align}
 t_n=t , \ \ \
 \alpha_n=\alpha, \ \ \ 
 \Delta_{Z,n}=\Delta_Z.
 \end{align}
In Figs.~\ref{figParameterProfileSmoothOneQDot} and \ref{figParameterProfileSmoothTwoQDot}, we show examples of profiles for the superconducting gap, the Zeeman energy, the chemical potential, the tunneling matrix element, and the Rashba SOI strength in an NS and an NSN junction. This system can enter a topological phase hosting MBSs, however, we will mainly focus on the trivial regime which can host only quasi-MBSs.

\section{ABS in Non-topological nanowires \label{SecNanowireSharpProfiles}}
 \subsection{ABS in the left  normal section }
 
In this section we study  ABSs in non-topological nanowires as defined in Sec. \ref{SubSec:NonTopSys}. 
We start our investigation with the setup shown in Fig.~\ref{figNanoWireScatteringRegionLeftNormalSectionBarrier} but without tunnel barriers or leads. 
The left (right) normal section can  host  ABSs localized  close to $\text{N}_1$ ($\text{N}_2$). The ratio $2 \alpha_{i} a/L_{i}$ determines the ABS level spacing and therefore the number of ABSs in the left ($i=1$) and right ($i=2$) normal sections.  If this ratio is large in comparison to $ \Delta_0 $ as is  in our case, then the system hosts only a few or a single ABS. The energy of the  ABS is pinned to zero if the parameters approximately fulfill the resonance condition 
\begin{align}
\cos(2k_{\text{so,i}}L_{i})=0 ,\label{eq:ResCond}
\end{align}
where $k_{\text{so,i}}=m_{i}\alpha'_{i}/\hbar^2$ denotes the SOI momentum and $ m_{i} = \hbar^2/(2t_{i} a^2) $ denotes the effective electron mass inside the normal section \cite{reeg2018zero}.  The resonance condition is derived for a chemical potential equal to zero in the normal section, where it is calculated from the SOI energy. The ABS energy, however, can also be  pinned to zero in the case of a non-zero chemical potential inside the normal section as we will demonstrate below numerically.  In particular, we discuss both the energy spectrum as well as wavefunctions of the ABS. The wavefunction contains  information about the spatial distribution of the ABS, which is important for understanding the local and non-local  differential conductance of ABSs in three-terminal devices.   
The parameter profiles for the NS junction are shown in Fig.~\ref{figParameterProfileSharpOneQDot}. 
We examine the case  of small ratios  $q(\Delta_Z)=L_S/\xi (\Delta_Z)$ between the length $L_S$ of the superconducting section and the localization length $\xi$ of the ABSs,
\begin{align}
\xi(\Delta_Z)=\hbar v_F/ \Delta. \label{eq:LocalizationLengthNonTopoABS}
\end{align}
Here, the renormalized Fermi velocity is defined as $v_F\approx \sqrt{2\mu_S/m_S}$ with $m_S=\hbar^2/(2t_Sa^2)$ being the effective mass in the superconducting section and the dependence of $\Delta$ on $\Delta_Z$   is defined in Eq. (\ref{de}).
The value of $q$ is small for a short superconducting section or for a small superconducting gap $ \Delta_0 $. The latter is associated with a large localization length since $ \xi $ is inverse proportional to $ \Delta $.

\begin{figure}[t]
\subfloat{\label{figEnergyOscillationResonance09DeltaZc} \stackinset{l}{-0.02in}{t}{0.0in}{(a)}{\stackinset{l}{1.7in}{t}{0.0in}{(b)}{\stackinset{l}{-0.02in}{t}{1.35in}{(c)}{\stackinset{l}{1.7in}{t}{1.35in}{(d)}{\includegraphics[width=1\columnwidth]{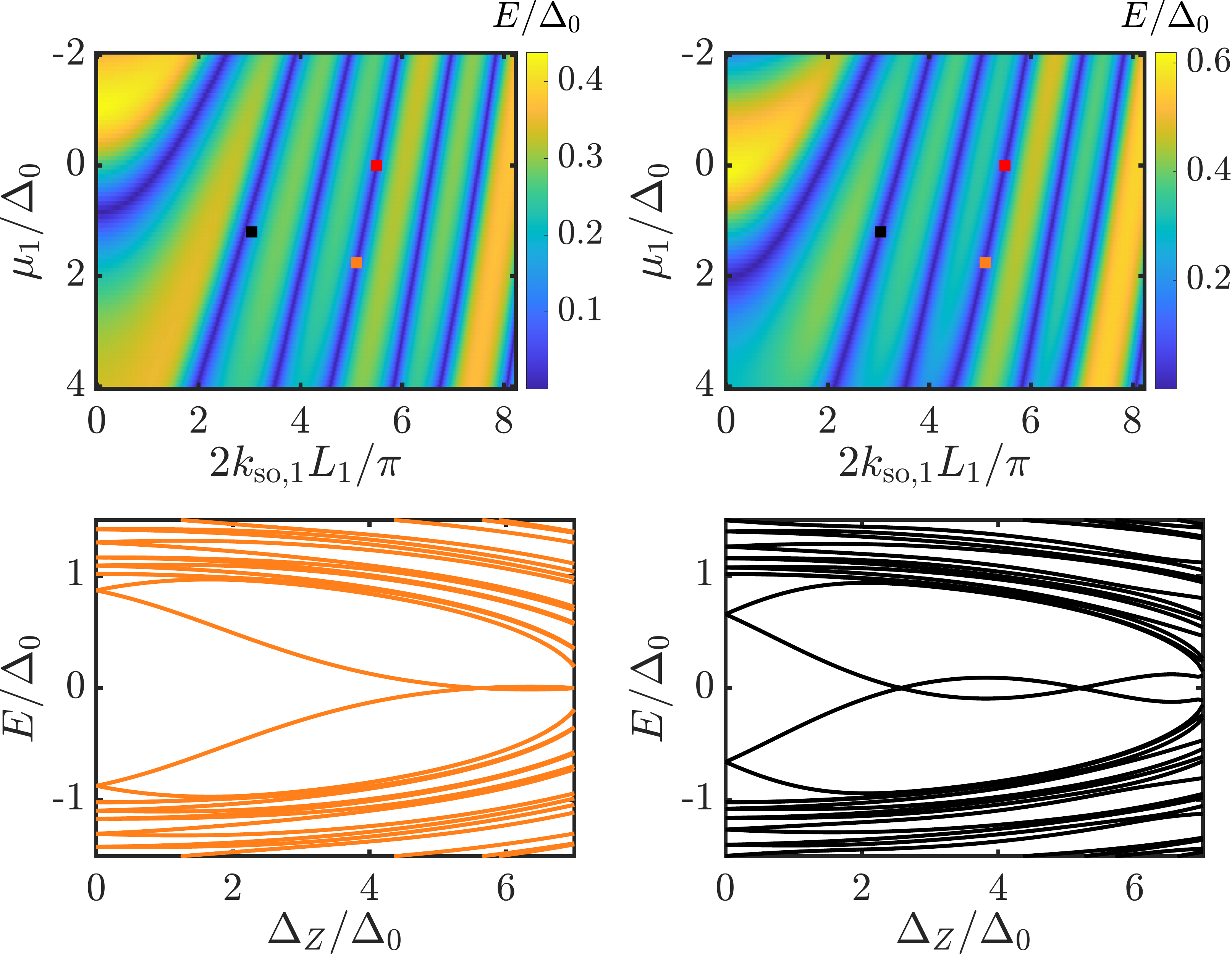}}}}}
}
\subfloat{\label{figEnergyOscillationResonance075DeltaZc}}
\subfloat{\label{figFiniteChemicalPotentialZeroBiasPinning}}
\subfloat{\label{figFiniteChemicalPotentialZeroBiasPinningProblemPhaseShift}}\\

\caption{Non-topological nanowire with an ABS in the left normal section [see Fig.~\ref{figNanoWireScatteringRegionLeftNormalSectionBarrier}]:  The energy of the  ABS oscillates as a function of the chemical potential $\mu_1$ and the SOI wavevector $k_{so}$ for the fixed Zeeman energy:  (a) $\Delta_Z=0.9\Delta_Z^c$    and  (b) $\Delta_Z=0.75\Delta_Z^c$. If the minima of the lowest  energy (blue regions), which determine the numerical resonance condition, appear at the same values of SOI strength and chemical potential (the red and the orange square) for the different Zeeman energies,   then the ABS stays at zero energy for some range of $\Delta_Z$, see panel  (c) and Fig.~\ref{figABSSingleQdotEnergySpectrumMagDependence1HugeGap}. Otherwise (the black square), the ABS energy is not strictly pinned to zero as shown in (d). These ABS levels are, however, broadened by finite temperature and by coupling to external leads in a transport experiment such that one can still observe an apparent ZBP, see App.~\ref{App:Broadening}.
The parameters are listed in Table~\ref{Tab:ParNonTopo1} in App.~\ref{App:Parameters}.}
 \label{figEnergyPinningAtNonZeroChemPot}
 \end{figure}
 
 \begin{figure}[t]
\subfloat{\label{figABSSingleQdotEnergySpectrumMagDependence1HugeGap} \stackinset{l}{0.0in}{t}{0.02in}{(a)}{\stackinset{l}{1.8in}{t}{0.02in}{(b)}{\stackinset{l}{-0.00in}{t}{1.17in}{(c)}{\stackinset{l}{1.8in}{t}{1.17in}{(d)}{\stackinset{l}{0in}{t}{2.34in}{(e)}{\stackinset{l}{1.8in}{t}{2.34in}{(f)}{\stackinset{l}{0in}{t}{3.51in}{(g)}{\stackinset{l}{1.8in}{t}{3.51in}{(h)}{\includegraphics[width=1\columnwidth]{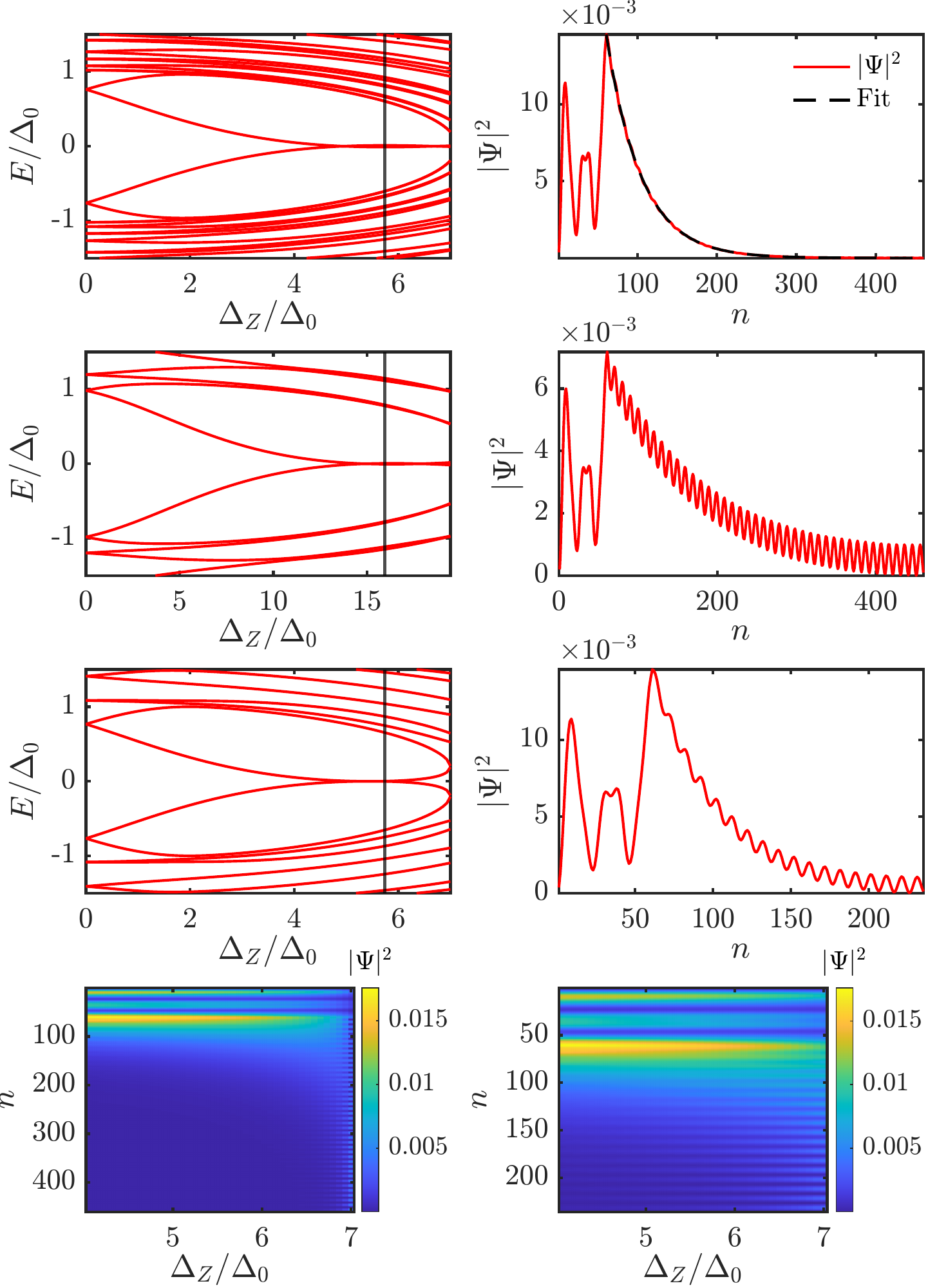}}}}}}}}}
}
\subfloat{ \label{figABSSingleQdotProbDensHugeGap} }
\subfloat{\label{figABSSingleQdotEnergySpectrumMagDependence1}}
\subfloat{ \label{figABSSingleQdotProbDensSmallDelta}}
\subfloat{\label{figABSSingleQdotEnergySpectrumMagDependence1ShortWire}}
\subfloat{ \label{figABSSingleQdotProbDensShortWire} }
\subfloat{\label{figSingleABSResoannceProbDensMagFieldDependencelongWirePinningRegion}}
\subfloat{\label{figSingleABSResoannceProbDensMagFieldDependenceshortWirePinningRegion}}
\caption{ (a,c,e)  Energy spectrum and (b,d,f) probability density of the ABS, $|\Psi|^2$,
 at $\Delta_{Z,0}=5.74\Delta_0$ (indicated by the black line in the left panels) for different values of $q=L_S/\xi$.
First row: In the case of  a long superconducting section $L_S\gg\xi$ [$q(\Delta_{Z,0})=4.5$], the ABS probability density is only non-zero on the left end of the nanowire and decays exponentially inside the superconducting section. We extract the numerical localization length of the ABS by fitting an exponential function (black dashed line) to the probability density. This numerically calculated localization length $ \xi=438$ nm agrees well with the analytic result  $ \xi=442$ nm.
Second (third) row corresponds to a small value of  $\Delta_0$ (of $L_S$) with $q(\Delta_{Z,0})=1.63$ [$q(\Delta_{Z,0})=1.98$]. In this case, $|\Psi|^2$ has a finite  value on the right end of the nanowire. Generally, as one approaches $\Delta_Z^c$, the ABS probability density also has a finite weight on the right end of the nanowire. For fixed $\Delta_0$, this effect is more pronounced in (h) short  than  in (g)  long nanowires. 
 The parameters are listed in Table~\ref{Tab:ParNonTopo1} in App.~\ref{App:Parameters}.
 \label{figSingleQDOTWaveFuncsEnergies}}
 \end{figure}

In Figs.~\ref{figEnergyOscillationResonance09DeltaZc} and \ref{figEnergyOscillationResonance075DeltaZc} we plot the energy of the lowest ABS as a function of the SOI momentum $k_{\text{so,1}}$ and of the chemical potential $\mu_1$ with the Zeeman energy being fixed close to $\Delta_Z^c$. For $\mu_1=0$ the ABS energy exhibits an oscillatory behavior that approximately matches  the resonance condition from Eq.~\eqref{eq:ResCond}. The oscillatory behavior  is preserved  for $\mu_1\neq 0$ and there are still recurring points at which the energy is close to zero (blue). Tuning the system to one of  these  resonance points also for finite values of the chemical potential (e.g. the orange or red square), we find a zero-energy pinning  (see Figs.~\ref{figFiniteChemicalPotentialZeroBiasPinning}  and \ref{figABSSingleQdotEnergySpectrumMagDependence1HugeGap}). If the resonance points do not  coincide for the different Zeeman energies (see the black square), then the energy is not  strictly pinned to zero, see Fig.~\ref{figFiniteChemicalPotentialZeroBiasPinningProblemPhaseShift}. In a transport experiment, however, such small deviations from zero energy could be masked by, for example, finite temperature, resulting in a broadened ZBP (see App.~\ref{App:Broadening}).

When the ratio  between the length of the superconducting section and the localization length  is large ($ q \gg 1$),
the exponential decay of the ABS wavefunction in the superconducting section means that the ABS is essentially entirely localized on the left side of the nanowire, see Fig.~\ref{figABSSingleQdotProbDensHugeGap}. We extract the localization length of the ABS from the numerically calculated probability density, see Fig.~\ref{figABSSingleQdotProbDensHugeGap},  and find that the numerical value of $\xi$ agrees well with the prediction of the analytic expression from Eq.~\eqref{eq:LocalizationLengthNonTopoABS}. 
A smaller $q$ can be achieved by choosing a smaller superconducting gap (see Figs.~\ref{figABSSingleQdotEnergySpectrumMagDependence1}  and \ref{figABSSingleQdotProbDensSmallDelta}) or decreasing the length of the superconducting section (see Figs.~\ref{figABSSingleQdotEnergySpectrumMagDependence1ShortWire} and \ref{figABSSingleQdotProbDensShortWire}). TAs the parameter $q$ approaches one, the exponential suppression becomes less pronounced.  his results in a small but finite probability density on the right end of the nanowire.  We note that the probability density of the ABS on the right side is always non-zero for large values of the Zeeman energy when the superconducting gap is suppressed, see Fig.~\ref{figSingleABSResoannceProbDensMagFieldDependenceshortWirePinningRegion}. This behavior is explained by the dependence of the localization length on the Zeeman energy. The localization length increases for large Zeeman energies and therefore the parameter $q$ approaches the value $q\sim1$. In addition, we note that the extended wavefunction of the ABS  in nanowires with small values of $q$ can be expected to generate a signature in the local  conductance measurements on both ends of the nanowire. 
These local signals  on  the left and right end would be correlated since they correspond to the same ABS. Experiments may therefore not be able to distinguish between this correlated ABS signatures and MBS signatures, when the parameter $q$ is small. 

\begin{figure}[t]
\subfloat{\label{figResonanceConductanceTemp50mKLeft}\stackinset{l}{-0.02in}{t}{0.02in}{(a)}{\stackinset{l}{1.75in}{t}{0.02in}{(b)}{\stackinset{l}{-0.02in}{t}{1.07in}{(c)}{\stackinset{l}{1.75in}{t}{1.07in}{(d)}{\stackinset{l}{-0.02in}{t}{2.1in}{(e)}{\stackinset{l}{1.75in}{t}{2.1in}{(f)}{\includegraphics[width=1\columnwidth]{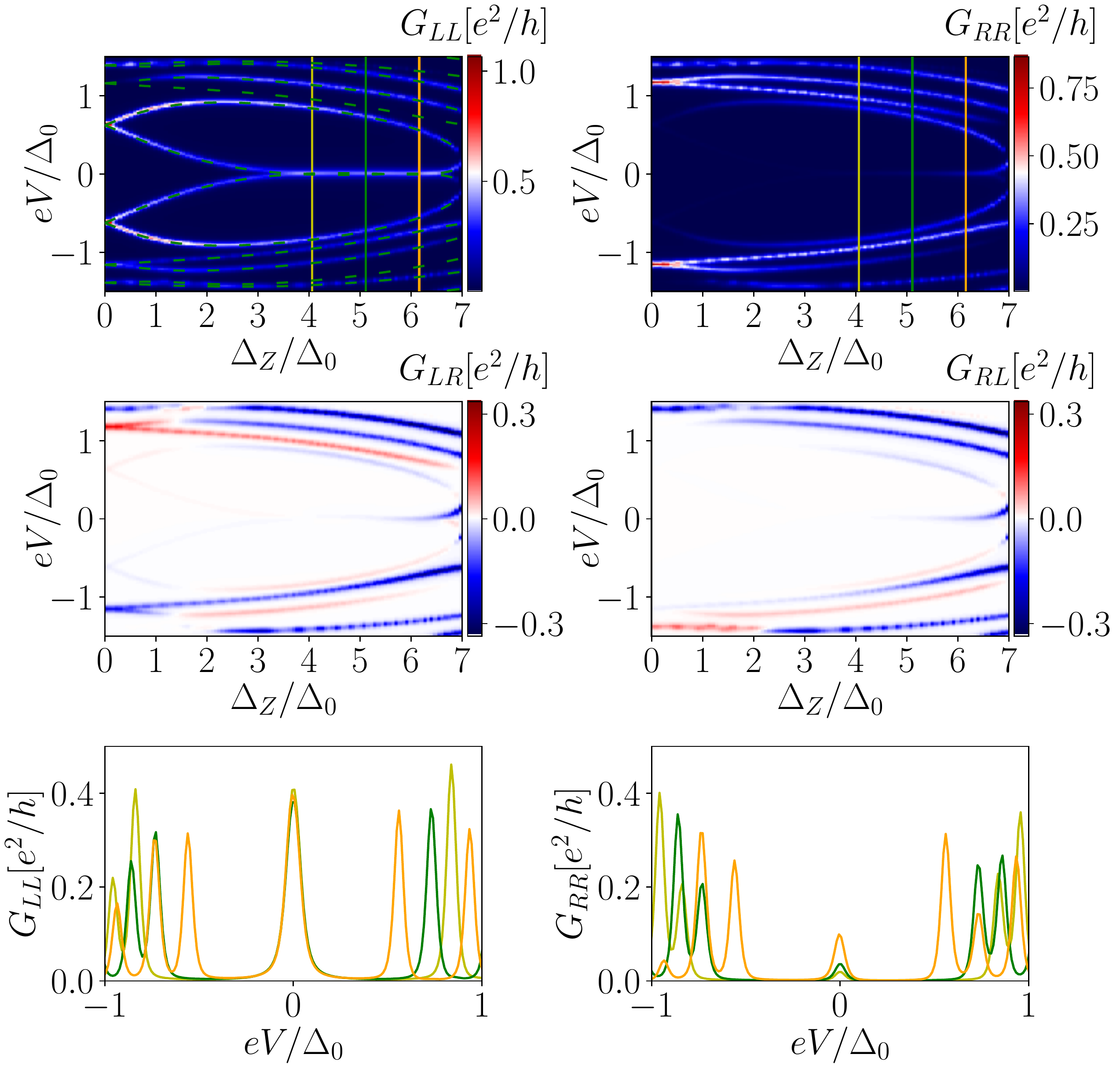}}}}}}}
}

\subfloat{\label{figResonanceConductanceTemp50mKRight}}
\subfloat{\label{figResonanceConductanceTemp50mKLeftRight}}
\subfloat{\label{figResonanceConductanceTemp50mKRightLeft}}
\subfloat{\label{figResonanceConductanceTemp50mKLeftLineCut}}
\subfloat{\label{figResonanceConductanceTemp50mKRightLineCut}}
  \caption{Differential conductance in a  non-topological nanowire containing one ABS on the left end that extends up to the right end. Both local  conductances (a) $G_{LL} $  and (b)  $G_{RR} $  exhibit a ZBP due to the extended nature of the ABS wavefunction.  The  conductance of the ABS is not quantized to $2e^2/h$ due to the shape of barriers chosen.  This conductance pattern agrees well with the energy spectrum, indicated by the dark green dashed lines. The yellow, dark green, and orange solid line indicate  linecuts of  (e)  $G_{LL} $ and (f) $G_{RR} $  at the Zeeman energies $\Delta_Z=\lbrace 4.01, 5.11, 6.16\rbrace \Delta_0$.  The non-local  conductances (c) $G_{LR} $ and (d) $G_{RL}$   contain signatures of the extended ABSs and of the bulk-gap closing at $\Delta_Z^c$.  The parameters are listed in Table~\ref{Tab:ParNonTopo1} in App.~\ref{App:Parameters}. }
 \label{figConductanceMatrixReegSystem1}
\end{figure}

Next, we calculate numerically the differential-conductance matrix elements $G_{\alpha\beta}=dI_{\alpha}/dV_{\beta}$, which are the derivative of the current $_{\alpha}$ in lead $ \alpha $ into the nanowire with respect to the voltage bias $V_{\beta}$ at lead $ \beta $   (we follow the notation of Ref.~\cite{DanonNonlocal2020}, see App.~\ref{App:DiffCond}). To account for tunnel barriers  and leads at both ends, see Fig.~\ref{figNanoWireScatteringRegionLeftNormalSectionBarrier}, we choose a slightly longer normal section $L_1$ than before. 
The local  conductance $G_{LL}$ on the left end exhibits very similar features as the energy spectrum, which we plot for comparison as dark green dashed lines, see Fig.~\ref{figResonanceConductanceTemp50mKLeft}. The ABS is visible for all Zeeman energies and is pinned close to zero  for a wide range of $\Delta_Z$ but the conductance is not quantized to $G=2e^2/h$ at zero bias and  depends on the tunnel barrier properties such as its strength and length, which would be also a case for MBSs.  Current experiments do not observe the 
quantized value, $2e^2/h$, of the ZBP expected for an MBS, thus, experiments cannot easily distinguish between this trivial feature and an MBS signature. A weaker ZBP also appears  in $ G_{RR} $ for $\Delta_Z\approx 4\Delta_0$ and stays stable until the superconducting gap is suppressed at $ \Delta_Z^c $, see also the line-cuts in Fig.~\ref{figResonanceConductanceTemp50mKRightLineCut}. This ZBP only appears for large  Zeeman energies when the wavefunction starts to leak  through the superconducting section. An equivalent signature could also be expected for the MBS case, for instance, when the two tunnel barriers are of different strength. 

The non-local   conductances $ G_{LR} $ and $G_{RL}$ are similar to each other and exhibit the bulk-gap closing at $ \Delta_Z=\Delta_Z^c $ as well as the ZBP, see Fig. \ref{figConductanceMatrixReegSystem1}. This ZBP in the non-local   conductance is not present in long nanowires but it is visible in short wires due to the extension of the ABS over the entire superconducting section. We note that non-zero non-local conductances indicate that the local  conductances $G_{LL}$ and $G_{RR} $ are not symmetric with respect to the bias, since electrons might tunnel directly between the normal leads, see Refs. \cite{DanonNonlocal2020, Melo2021Conductance}.  The sum of all differential-conductance matrix elements, however, is symmetric with respect to the bias. The antisymmetric part of the local conductance $G_{LL}^a$ ($G_{RR}^a$) corresponds to the negative value of the antisymmetric part of the non-local  conductance $G_{LR}^a$ ($G_{RL}^a$), see  Ref. \cite{DanonNonlocal2020}.

The ZBP in our setup is robust against changes of the Zeeman energy but not against fluctuations of the tunnel barrier strength $ \gamma_1 $. Indeed, tuning $ \gamma_1 $ to slightly different values removes the perfect zero-energy pinning. Parenthetically we note that in short topological nanowires,  the MBS wavefunctions overlap, and so, similar to the behavior of our ABSs, it is anyway expected that MBSs are not fixed to zero energy in short wires. Furthermore, broadening effects, for example due to temperature, affect the differential conductance. If the energy is not perfectly pinned to zero and the broadening is large enough then a conductance measurement can not resolve a small finite energy splitting and will reveal only a single peak, which actually consists of two single merged peaks around zero bias, see App.~\ref{App:Broadening}. Although our system is not designed to explain the data from any specific experiment, we note that our results are similar to the experimental data from Ref. \cite{yu2020non}. In particular, a ZBP appears in the left conductance for a specific value of the tunnel barrier gate voltage whereas a ZBP appears in the right  conductance at larger Zeeman energies. 

We conclude that such an ABS mimics certain key properties of an MBS, which, in turn, presents a challenge for an unambiguous interpretation of experimental observations. If the ratio between the length of the  superconducting section and the localization length is small, then the ABS-ZBPs can even be correlated at  the left and right ends of the nanowire.  The ABS requires some tuning and is not universally stable against fluctuations in the SOI strength or the tunnel barrier strength. We again note that, by construction, the system considered in this section cannot enter the topological phase and so all features we have found here are due to trivial ABSs.

\subsection{ABS in the left and right normal sections \label{SubSecResonanceNSNjunction}}
\begin{figure}[t]
\subfloat{\label{figABSTwoQdotEnergySpectrumMagDependenceBothInResonance} \stackinset{l}{0.0in}{t}{0.02in}{(a)}{\stackinset{l}{1.7in}{t}{0.02in}{(b)}{\stackinset{l}{-0.00in}{t}{1.17in}{(c)}{\stackinset{l}{1.7in}{t}{1.17in}{(d)}{\stackinset{l}{0in}{t}{2.34in}{(e)}{\stackinset{l}{1.7in}{t}{2.34in}{(f)}{\stackinset{l}{0in}{t}{3.51in}{(g)}{\stackinset{l}{1.7in}{t}{3.51in}{(h)}{\includegraphics[width=1\columnwidth]{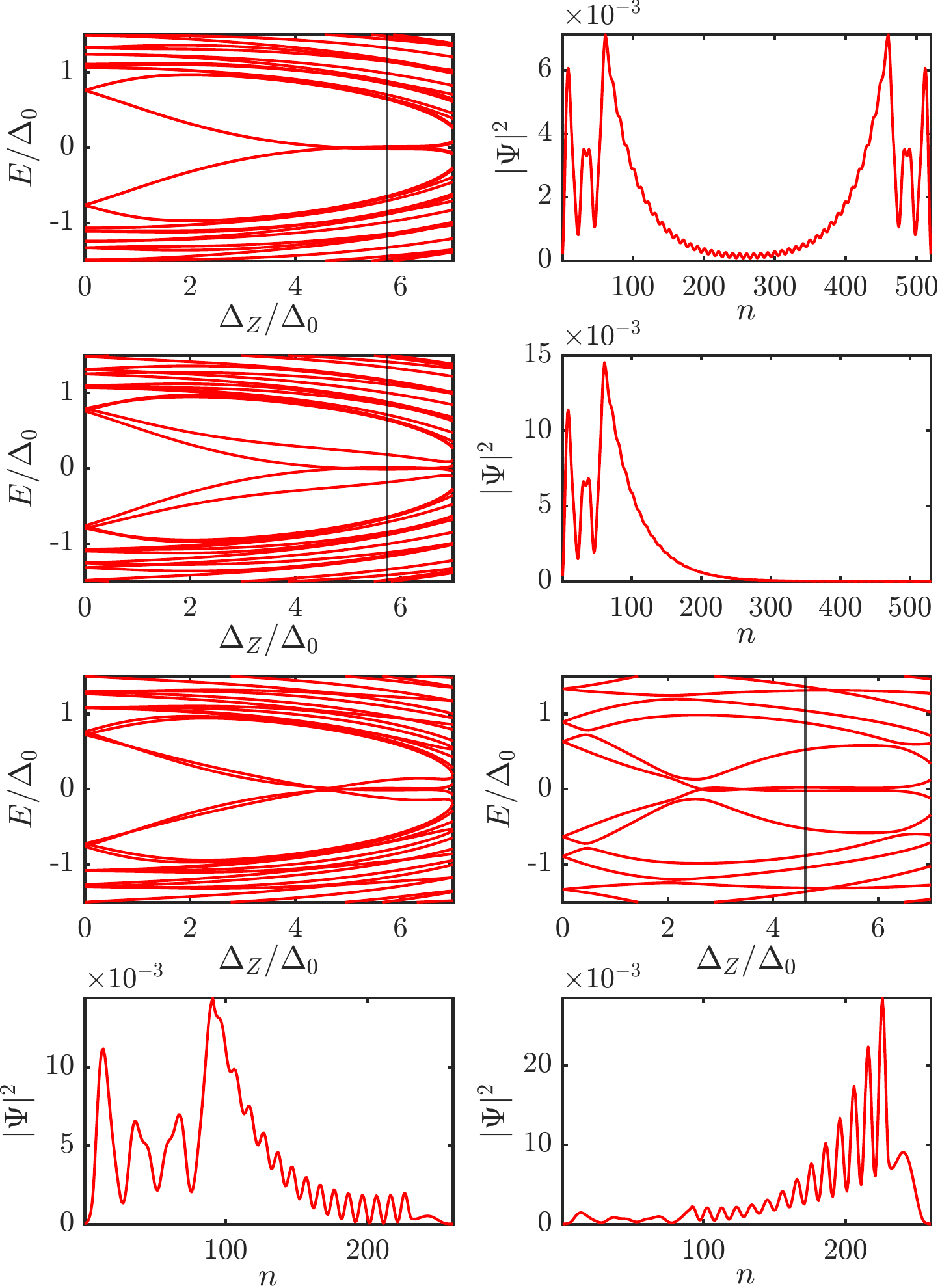}}}}}}}}}
}
\subfloat{\label{figABSTwoQdotProbDensBothInResonance}}
\subfloat{\label{figABSTwoQdotEnergySpectrumMagDependenceOneInResonance}}
\subfloat{\label{figABSTwoQdotProbDensOneInResonance}}
\subfloat{\label{figTwoQDotsOneABSMimicsGapclosingLongtWire2}}
\subfloat{\label{figTwoQDotsOneABSMimicsGapclosingshortWire4}}
\subfloat{\label{figTwoQDotsShortWireAntiCrossingABSLeft4}}
\subfloat{\label{figTwoQDotsShortWireAntiCrossingABSRight4}}
\caption{Non-topological nanowire with two ABSs: one ABS in the left and a second one in the right normal section. (a,c,e,f) Energy spectrum and  (b,d,g,h) probability densities of the ABSs at $\Delta_Z=5.74\Delta_0$ and $\Delta_Z=4.62\Delta_0$ (indicated by the black line in the panels with the energy spectra). First row: Both ABSs in $\text{N}_1$ and $\text{N}_2$ are tuned to zero energy, therefore the ABSs in the left and right normal sections are degenerate but essentially uncorrelated. Tuning the parameters of the right ABS away from the resonance condition (second row), one lifts the degeneracy. The left ABS stays pinned to zero energy, see (d).
 Third row: The right ABS mimics the behaviour of the edge of a bulk-gap in (e)  long and  (f) short nanowires.
The probability density of the (g) left  and (h)  right ABS, corresponding to the panel (f), has a finite value throughout the entire nanowire. The parameters are listed in Table~\ref{Tab:ParNonTopo1} in App.~\ref{App:Parameters}. }
\label{figTwoQDOTWaveFuncsEnergies}
\end{figure}

In this section, we examine the non-topological nanowire with two normal sections hosting two ABSs: one on the left and another one on the right side of the nanowire, see Fig.~\ref{figNanoWire1}. As before we begin without tunnel barriers and without the leads.
If the resonance condition is fulfilled simultaneously in the left and the right normal sections of the long nanowire, then the two ABSs  become degenerate. The probability density shows  peaks at both ends of the nanowire, see Fig.~\ref{figABSTwoQdotProbDensBothInResonance}.  
For long wires there is no correlation between the ABS on the left end  and the ABS on the right end: both are independent of each other and the overlap of their wavefunctions is approximately zero. As can be expected from the previous section, this is not the case for shorter wires and correlations can occur when the ratio $q$ is small.

In general, a topological phase transition is accompanied by a bulk-gap closing and reopening. Here, we show that such a gap behavior can also be mimicked by  two ABSs in non-topological nanowires.
We tune the parameters of the right normal section away from the resonance condition by changing the length of $\text{N}_2$. The degeneracy is lifted and the energy of the right ABS is different from that of the left ABS, see Fig.~\ref{figABSTwoQdotEnergySpectrumMagDependenceOneInResonance}. The parameters $\alpha_2$ and $N_2$ do not affect the zero-energy pinning of the left ABS and can be chosen independently to control the behavior of the right ABS in dependence of the  Zeeman energy. We then tune the right ABS such that it crosses the zero energy at the same value  of the magnetic field at which the zero-energy pinning of the left ABS starts to take place.  The resulting energy spectrum is shown in Fig.~\ref{figTwoQDotsOneABSMimicsGapclosingLongtWire2} and is reminiscent of what one might expect close to the topological phase transition, however, we stress that here all these features occur due to the presence of trivial ABSs in non-topological nanowires.

The  nanowire examined in Fig.~\ref{figTwoQDotsOneABSMimicsGapclosingLongtWire2} is relatively long with a large value of the parameter $q\approx4.5$, it is therefore not expected that the left ABS is visible in the local  conductance on the right end of the nanowire. If instead we choose a similar parameter set as in Fig.~\ref{figConductanceMatrixReegSystem1}, corresponding to a short nanowire and, in addition, account for a tunnel barrier  (see  Fig.~\ref{figNanoWire1}), we find the energy spectrum shown in Fig.~\ref{figTwoQDotsOneABSMimicsGapclosingshortWire4}. The energy spectrum in Fig.~\ref{figTwoQDotsOneABSMimicsGapclosingshortWire4} strongly resembles the gap closing and reopening one expects from a topological phase transition, but is again entirely due to ABSs.
Additionally, the wavefunction of the left ABS now spreads from the left to the right  end and vice versa for the wavefunction of the right ABS,
see Figs.~\ref{figTwoQDotsShortWireAntiCrossingABSLeft4} and \ref{figTwoQDotsShortWireAntiCrossingABSRight4}. 

   \begin{figure}[t]
\subfloat{\label{figResonanceConductanceTemp40mKLeft2}\stackinset{l}{-0.02in}{t}{0.02in}{(a)}{\stackinset{l}{1.7in}{t}{0.02in}{(b)}{\stackinset{l}{-0.02in}{t}{1.17in}{(c)}{\stackinset{l}{1.7in}{t}{1.17in}{(d)}{\includegraphics[width=1\columnwidth]{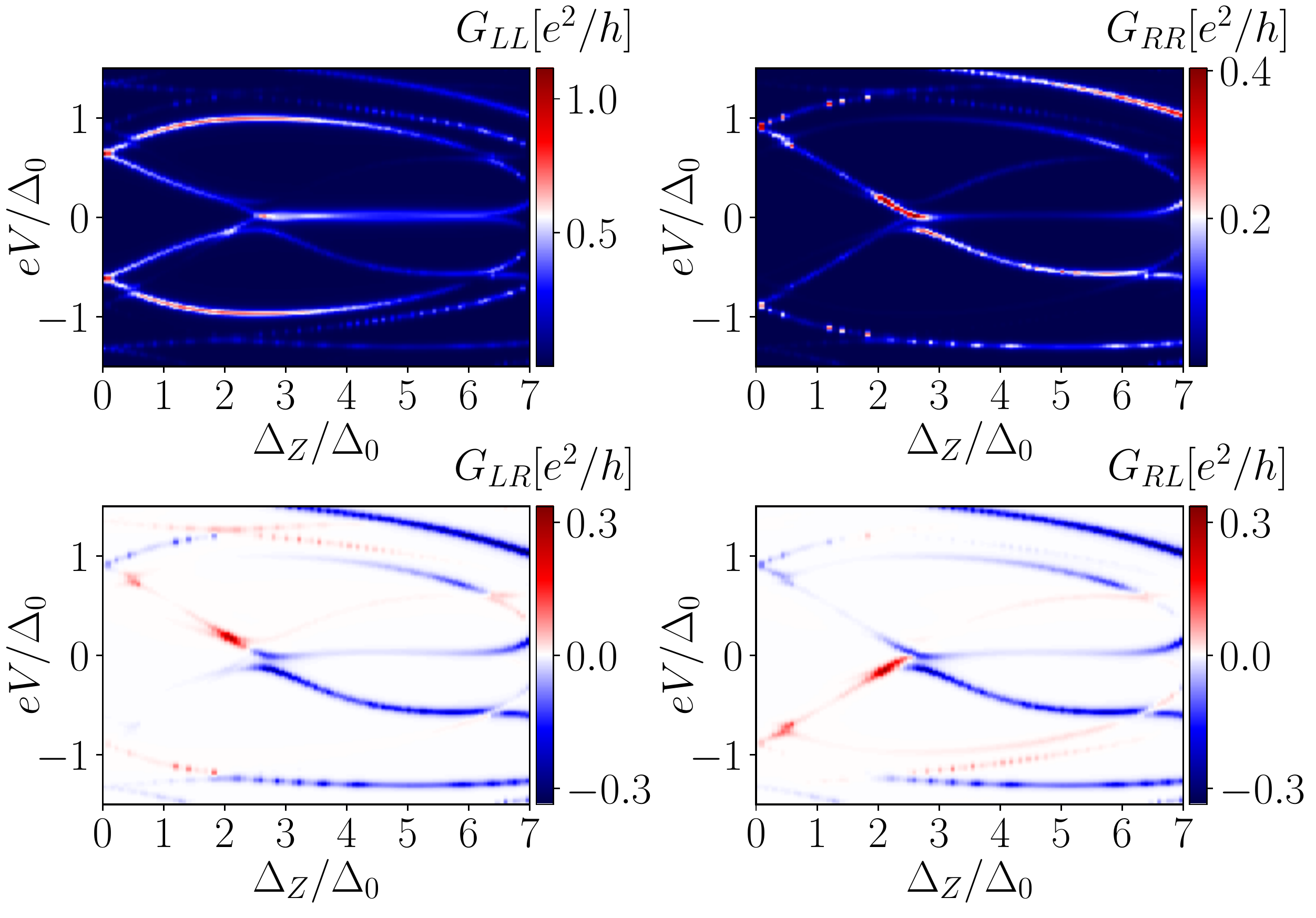}}}}}
}
\subfloat{\label{figResonanceConductanceTemp40mKRight2}}
\subfloat{\label{figResonanceConductanceTemp40mKLeftRight2}}
\subfloat{\label{figResonanceConductanceTemp40mKRightLeft2}}
 \caption{Differential-conductance  patterns corresponding to the energy spectrum of a non-topological nanowire from Fig.~\ref{figTwoQDotsOneABSMimicsGapclosingshortWire4}.   Both local conductances (a) $G_{LL} $   and (b) $G_{RR} $  exhibit a ZBP due to the extended left ABS wavefunction,  see Fig.~\ref{figTwoQDotsShortWireAntiCrossingABSLeft4}.  Although entirely trivial in origin, the local  conductance is reminiscent of what is expected for MBSs, containing both correlated ZBPs and an apparent gap closing and reopening process. The non-local  conductances (c) $G_{LR} $  and (d) $G_{RL}$  contain signatures of both  the bulk states and the extended ABSs, which are  similar to those expected for MBSs. The parameters are listed in Table~\ref{Tab:ParNonTopo1} in App.~\ref{App:Parameters}.  }
 \label{figConductanceMatrixReegSystemTwoABSMimicTPT3}
\end{figure}

The local  conductances $G_{LL}$ and $G_{RR}$ reveal that the ABS localized more on the left (right) is still visible at the opposite right (left) end, see Fig.~\ref{figConductanceMatrixReegSystemTwoABSMimicTPT3}. The left (right) ABS has a smaller conductance value on the right  (left) end and the  conductance is not quantized. In the absence of quantized conductances, however, this behavior significantly complicates the interpretation of future experimental data: The local  conductance on the left and right end exhibits a correlated ZBP and this is accompanied by a signature reminiscent of a bulk-gap closing and reopening. In addition, the non-local  conductance also exhibits the correlated left and right ABS-ZBP as well as a signature similar to a bulk-gap closing and reopening during a topological phase transition. All these features could be misinterpreted as signatures of MBSs but appear here in a nanowire that is, by design, topologically trivial at all magnetic field strengths. The complimentary scenario in short topological nanowires is discussed in App.~\ref{App:TopoNanowireAdditionalMaterial}.

\begin{figure}[t]
\subfloat{\label{figLongWireSmoothGapandPotEnergies}\stackinset{l}{-0.03in}{t}{0.05in}{(a)}{\stackinset{l}{1.55in}{t}{0.05in}{(b)}{\includegraphics[width=1\columnwidth]{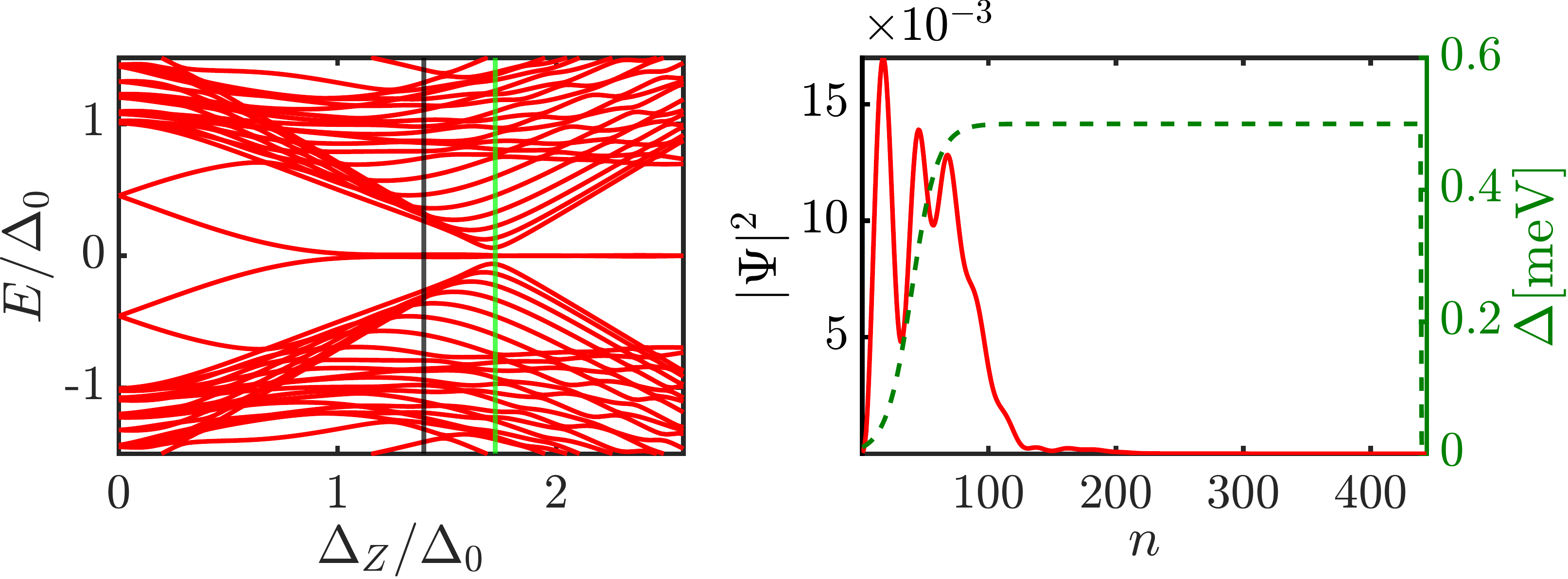}}}
}
\subfloat{\label{figLongWireSmoothGapandPotProbDens}}
\vspace{-0.2in}
\subfloat{\label{figShortWireSmoothGapandPotEnergies}\stackinset{l}{-0.03in}{t}{0.05in}{(c)}{\stackinset{l}{1.55in}{t}{0.05in}{(d)}{\includegraphics[width=1\columnwidth]{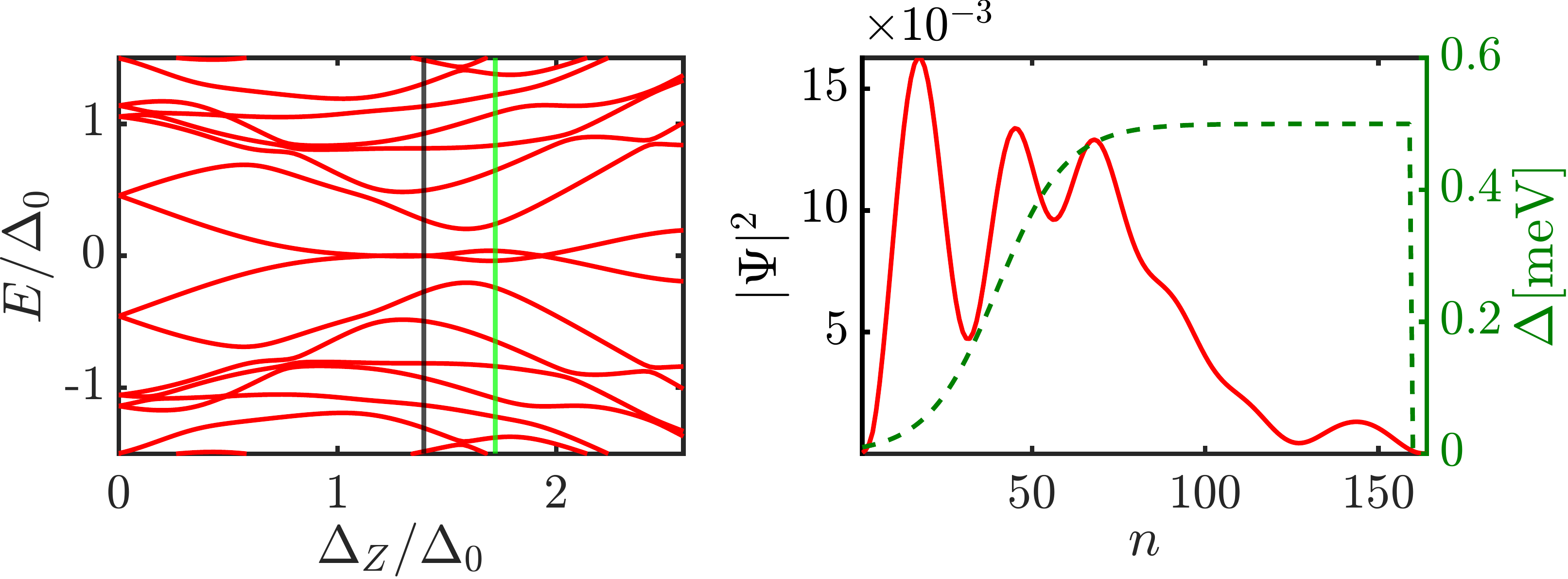}}}
}
\subfloat{\label{figShortWireSmoothGapandPotProbDens}}
 \caption{Topological nanowire with quasi-MBSs on the left end: (a,c) Energy spectrum and (b,d) probability densities of the quasi-MBS at $\Delta_Z=1.39\Delta_0$ (indicated by the black line in the left panels) and profile of the superconducting gap (dark green dashed line).  First row: In the case of a long superconducting section $L_S\gg \xi$, the quasi-MBS probability density is only non-zero on the left end of the nanowire and decays  inside the superconducting section. The second row corresponds to a smaller value of $L_S$ and in this case the probability density has a finite value at the right end of the nanowire. The topological phase transition from quasi-MBSs to MBSs takes place at $\Delta_Z=1.73\Delta_0$ (indicated by the green line in the left panels).  The parameters are listed in Table~\ref{Tab:ParTopo} in App.~\ref{App:Parameters}. }
 \label{figSmoothGapSmoothWireTPTEnergiesWavefunctions}
 \end{figure}

 \section{quasi-MBS in topological nanowires} \label{sec:NanoWireSmooth}
 \subsection{Quasi-MBS  in the left  normal section}
In this section, we consider topological nanowires in configurations shown in Fig.~\ref{figNanoWireScatteringRegionLeftNormalSectionBarrier} with parameter profiles shown in Fig.~\ref{figParameterProfileSmoothOneQDot}. Such nanowires host quasi-MBSs even if the superconducting section is in  the trivial phase  as discussed in Sec.~\ref{SubSec:TopSys}. 
In Fig.~\ref{figSmoothGapSmoothWireTPTEnergiesWavefunctions}, we compare the energy spectrum and probability density of systems with long and short superconducting sections.
Quasi-MBSs at approximately zero energy exist in the trivial phase and evolve into MBSs at stronger magnetic fields. The phase transition takes place approximately  at the critical value $\Delta_Z^T=\sqrt{\Delta_0^2+\mu_S^2}$, indicated by the green line, and is accompanied by a bulk-gap closing and reopening. Changing the shape of $ \Delta_n$ and $\mu_n$ to step-like functions shifts quasi-MBSs to higher energies, whereas MBSs in the topological phase are not affected. The wavefunctions of the quasi-MBSs only have support on the left end of the nanowire and decay inside the superconducting section. Therefore, the probability density is only non-zero also on the right end of the nanowire when MBSs appear. The quasi-MBSs still exist in short nanowires with a small ratio $q$ and, in this case, the wavefunction spreads through the superconducting section to the right end, see Fig.~\ref{figShortWireSmoothGapandPotProbDens}. In contrast to ABSs in the non-topological nanowire system considered above [see  Sec.~\ref{SecNanowireSharpProfiles}], quasi-MBSs in a nanowire with smooth parameter profiles are more  stable against fluctuations of the tunnel-barrier strength. For long wires quasi-MBSs can appear over a wide range of SOI strengths. In short nanowires, however, the quasi-MBSs are only pinned to zero for a narrow interval of the SOI strength.

\begin{figure}[t]
\subfloat{\label{figShortSmoothGapANDPotConductanceTemp50mKLeft4}\stackinset{l}{-0.00in}{t}{-0.0in}{(a)}{\stackinset{l}{1.67in}{t}{-0.0in}{(b)}{\stackinset{l}{-0.0in}{t}{1.08in}{(c)}{\stackinset{l}{1.67in}{t}{1.08in}{(d)}{\stackinset{l}{-0.0in}{t}{2.16in}{(e)}{\stackinset{l}{1.67in}{t}{2.16in}{(f)}{\includegraphics[width=1\columnwidth]{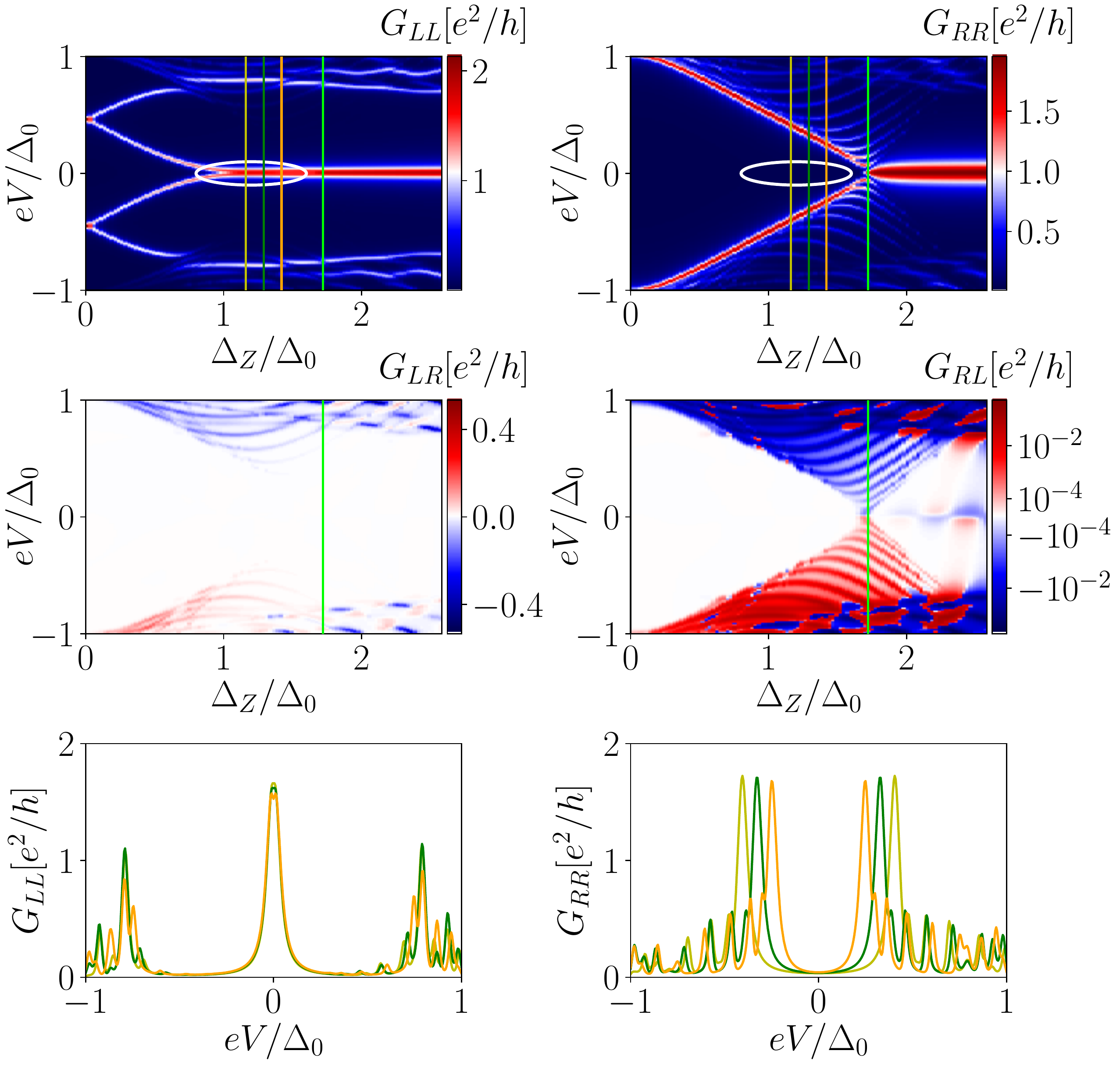}}}}}}}}
\subfloat{\label{figShortmoothGapConductanceTemp50mKRight4}}
\subfloat{\label{figShortSmoothGapConductanceTemp50mKLeftRight4}}
\subfloat{\label{figShortSmoothGapConductanceTemp50mKRightLeft4}}
\subfloat{\label{figLongWireLineCutGLL}}
\subfloat{\label{figLongWireLineCutGRR}}
  \caption{Differential-conductance patterns reproduce the energy spectrum of the topological nanowire, see Fig.~\ref{figLongWireSmoothGapandPotEnergies}.    The system undergoes a topological phase transition at $ \Delta_Z\approx1.73\Delta_0  $ as indicated by the green line.  The  local  conductance (a) $G_{LL} $   and  (b)   $G_{RR} $ of the  MBSs are  nearly quantized close to the value $2e^2/h$. The parameter region of potential quasi-MBSs is encircled by the white ellipse. Only $G_{LL} $  exhibits a ZBP of quasi-MBS: this local  conductance  is also quantized close to $2e^2/h$, deviations from this value are due to thermal broadening. The non-local  conductances  (c) $G_{LR} $  and (d)  $G_{RL}$ contain only signatures coming from the bulk states. Line cuts (e,f) of the local  conductance  $G_{LL} $ and $G_{RR} $ at the Zeeman energies $\Delta_Z=\lbrace1.16, 1.29, 1.42\rbrace\Delta_0$  [indicated by the yellow, dark green and orange lines in (a,b)] confirm that the quasi-MBS-ZBP appears only on the left end. The parameters are listed in Table~\ref{Tab:ParTopo} in App.~\ref{App:Parameters}. }
 \label{figConductanceMatrixSmoothSystemShortWire4}
\end{figure}

 \begin{figure}[t]
\subfloat{\label{figShortSmoothGapANDPotConductanceTemp50mKLeft3}\stackinset{l}{-0.00in}{t}{-0.0in}{(a)}{\stackinset{l}{1.67in}{t}{-0.0in}{(b)}{\stackinset{l}{-0.0in}{t}{1.08in}{(c)}{\stackinset{l}{1.67in}{t}{1.08in}{(d)}{\stackinset{l}{-0.0in}{t}{2.16in}{(e)}{\stackinset{l}{1.67in}{t}{2.16in}{(f)}{\includegraphics[width=1.0\columnwidth]{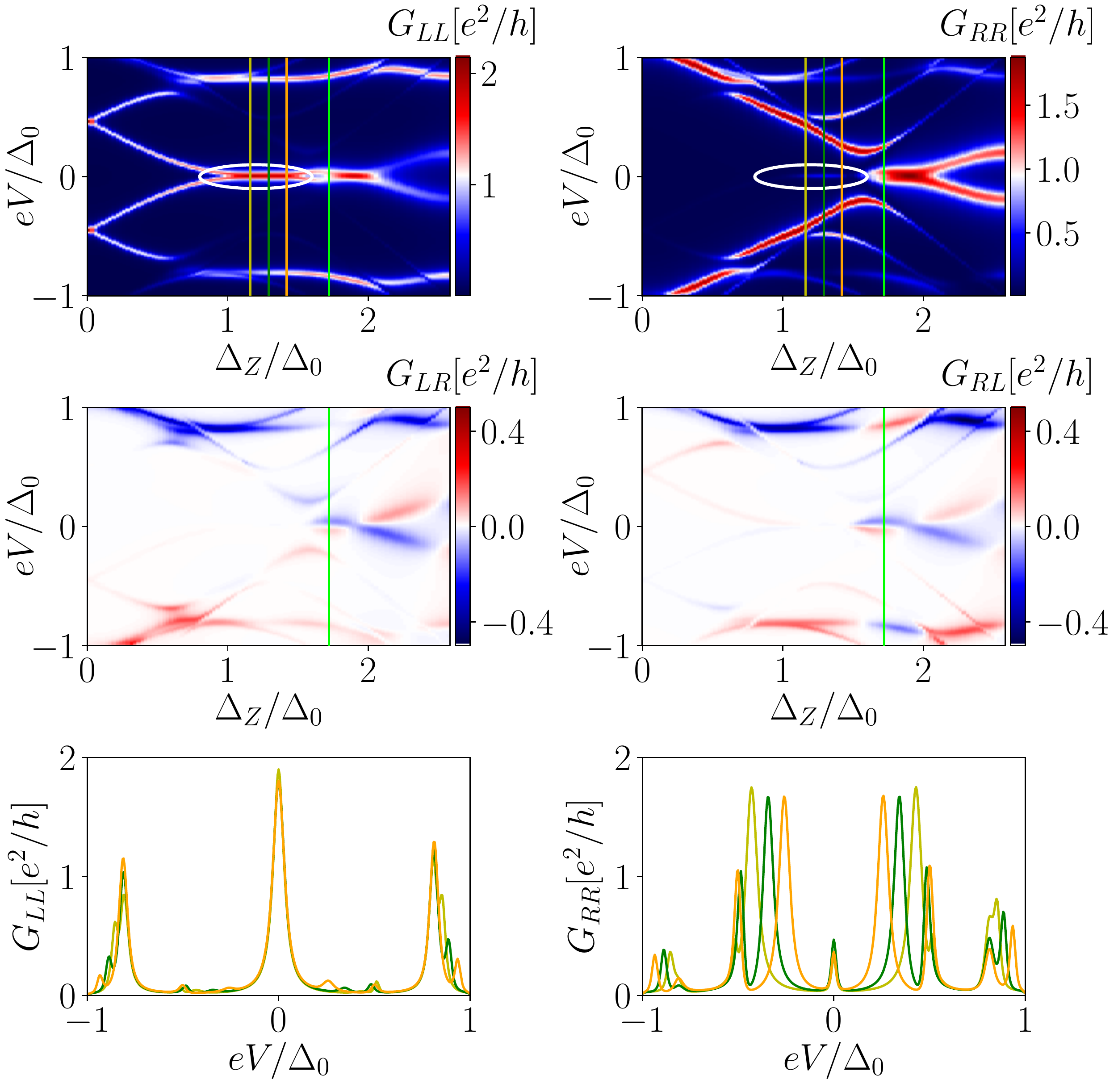}}}}}}}
}
\subfloat{\label{figShortmoothGapConductanceTemp50mKRight3}}
\subfloat{\label{figShortSmoothGapConductanceTemp50mKLeftRight3}}
\subfloat{\label{figShortSmoothGapConductanceTemp50mKRightLeft3}}
\subfloat{\label{figShortWireLineCutGLL}}
\subfloat{\label{figShortWireLineCutGRR}}
 \caption{The same as in  Fig.~\ref{figConductanceMatrixSmoothSystemShortWire4} but for a short nanowire. The  corresponding energy spectrum is shown in Fig.~\ref{figShortWireSmoothGapandPotEnergies}. Both local conductances  (a)  $G_{LL} $ and  (b) $G_{RR} $  exhibit a ZBP coming from quasi-MBSs  (highlighted by the white ellipse). The conductance of the quasi-MBSs is close to the quantization value of $2e^2/h$ on the left end but not on the right end. The non-local conductances  (c) $G_{LR} $  and  (d) $G_{RL}$ contain signatures of the bulk states as well as of the quasi-MBSs and MBSs. 
Line cuts (e,f) of the local  conductance  $G_{LL} $ and $G_{RR} $ at the Zeeman energies $\Delta_Z=\lbrace1.16, 1.29, 1.42\rbrace\Delta_0$
[indicated by the yellow, dark green, and orange lines in (a,b)] confirm  that the quasi-MBS-ZBP appears on both ends of the nanowire. The parameters are listed in Table~\ref{Tab:ParTopo} in App.~\ref{App:Parameters}.}
 \label{figShortWireSmoothParameterConductance}
\end{figure}

Within this setup we first study the transport properties of long topological nanowires that host quasi-MBSs, see Fig.~\ref{figConductanceMatrixSmoothSystemShortWire4}. As is expected for MBSs, the conductance of these quasi-MBSs is nearly quantized  to $2e^2/h$ for some set of parameters, as discussed in earlier works \cite{Penaranda2018Quantifying}. Deviations from this value are due to line broadening effects.  In long nanowires quasi-MBSs are only visible in the local  conductance on the left end, $G_{LL} $, the corresponding region is encircled by an ellipse in Figs.~\ref{figShortSmoothGapANDPotConductanceTemp50mKLeft4} and \ref{figShortmoothGapConductanceTemp50mKRight4}; see  also Figs.~\ref{figLongWireLineCutGLL} and \ref{figLongWireLineCutGRR} for line-cuts of the local conductances $G_{LL} $ and $G_{RR} $ at certain Zeeman energies. This behavior can be understood from the fact that the quasi-MBS wavefunction is  localized on the left end of the nanowire. The bulk-gap  closing and reopening is only weakly pronounced in  $G_{LL} $ 
because the bulk states are mainly localized within the superconducting section and the left lead is relative far away from this region. As a result, 
 $G_{LL} $ 
primarily probes the quasi-MBS (which is localized in $\text{N}_1$) but not the bulk states. It should be noted that the bulk states can become more visible using a logarithmic color scale (see Fig.~ \ref{figShortSmoothGapConductanceTemp50mKRightLeft4}). The normal section on the right end is shorter and so the right local conductance is a better probe of the bulk states. The bulk-gap closing and reopening in the non-local conductances  $G_{LR} $ and  $G_{RL} $, shown in Figs.~\ref{figShortSmoothGapConductanceTemp50mKLeftRight4} and \ref{figShortSmoothGapConductanceTemp50mKRightLeft4}, respectively, is less clear compared to nanowires with uniform parameters as well as the quasi-MBSs and the MBSs are not visible in the non-local  conductances. The right local conductance, $G_{RR}$, takes larger values close to the bulk-gap edge in the trivial regime (see Fig.~\ref{figShortmoothGapConductanceTemp50mKRight4}) since there is an `intrinsic' ABS just at the gap edge, see Refs. \cite{Huang2018Meta,Aseev2018Lifetime}. 

Next, we consider short superconducting sections. The conductance value of the quasi-MBS and its zero-bias pinning is essentially unaffected by the change of length, see Fig.~\ref{figShortWireSmoothParameterConductance}. In contrast, the MBSs that occur in the topological phase are pushed away from zero energy. In short nanowires quasi-MBSs are visible in $G_{RR} $:
this region is indicated by the white ellipse in Figs.~\ref{figShortSmoothGapANDPotConductanceTemp50mKLeft3} and \ref{figShortmoothGapConductanceTemp50mKRight3}. The quasi-MBS-ZBP appearing in $G_{RR} $
is not quantized and much smaller than that in $G_{LL} $,
see also Figs.~\ref{figShortWireLineCutGLL} and \ref{figShortWireLineCutGRR} for a line cut of the conductance.  The right local  conductance, however, exhibits a small ZBP and this peak is correlated to the one on the left end. Furthermore, while the quasi-MBSs and the MBSs generate a signal in  the non-local  conductances, $G_{LR} $ and  $G_{RL} $, the bulk-gap closing and reopening is not as clear as in the case of the long superconducting section.

 \subsection{Quasi-MBSs in the left and right normal sections}

 \begin{figure}[t]
\subfloat{\label{figLongSmoothGapConductanceTemp40mKLeft5}\stackinset{l}{-0.00in}{t}{-0.0in}{(a)}{\stackinset{l}{1.67in}{t}{-0.0in}{(b)}{\stackinset{l}{-0.0in}{t}{1.08in}{(c)}{\stackinset{l}{1.67in}{t}{1.08in}{(d)}{\includegraphics[width=1\columnwidth]{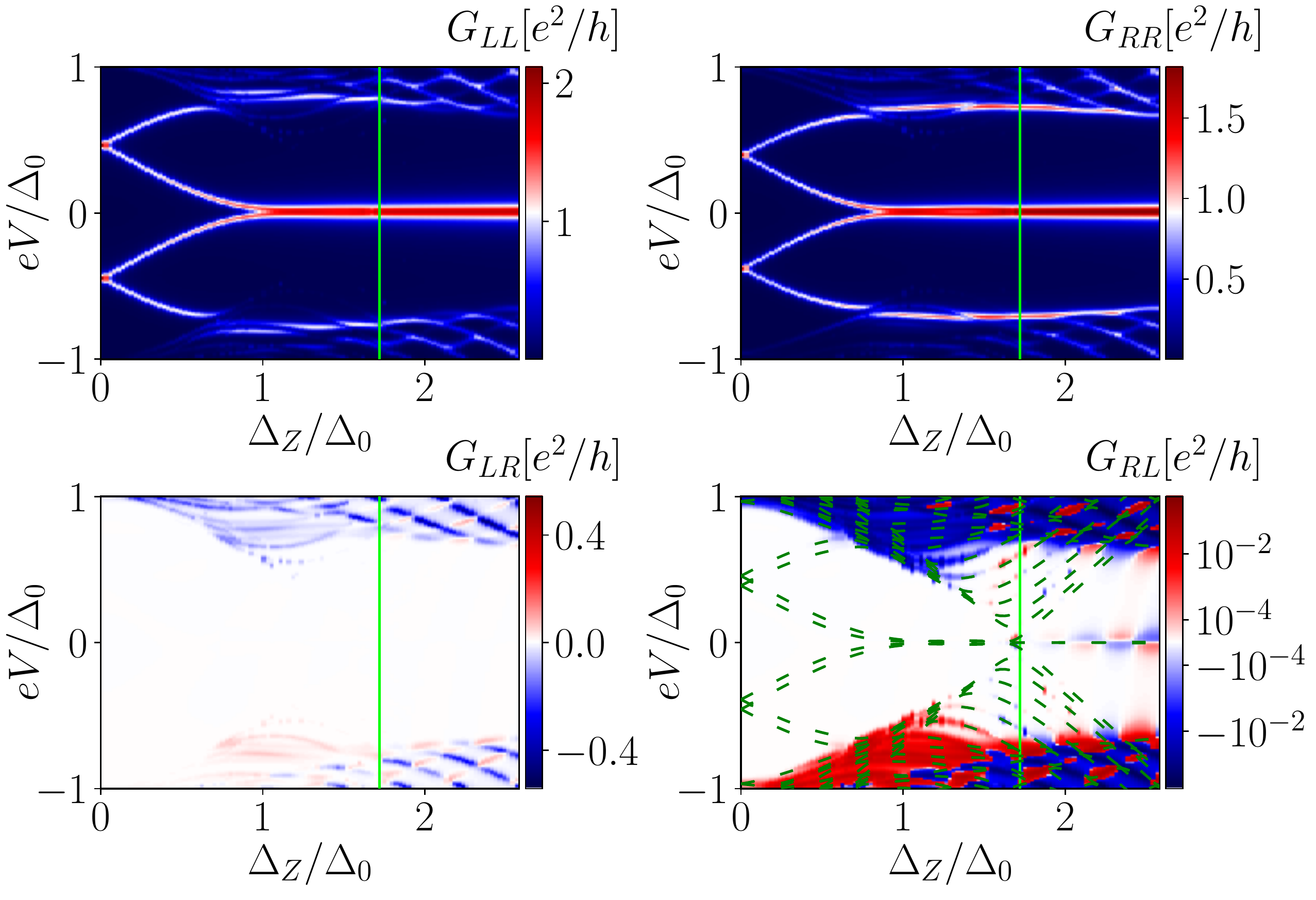}}}}}
}
\subfloat{\label{figLongSmoothGapConductanceTemp40mKRight5}}
\subfloat{\label{figLongSmoothGapConductanceTemp40mKLeftRight5}}
\subfloat{\label{figShortSmoothGapConductanceTemp40mKRightLeft5}}
  \caption{A long topological nanowire, as in Fig.~\ref{figShortSmoothGapANDPotConductanceTemp50mKLeft4}, with quasi-MBSs present at both nanowire ends.   The  local  conductance (a) $G_{LL} $   and  (b) $G_{RR} $  of the  MBSs and the quasi-MBSs is close to the quantization value of $G=2e^2/h$, deviations from this value are due to thermal broadening. The non-local  conductances (c) $G_{LR} $  and (d) $G_{RL}$  contain  only signatures of the bulk states, the bulk-gap closing and reopening is only weakly pronounced.  The parameters are listed in Table~\ref{Tab:ParTopo} in App.~\ref{App:Parameters}. }
 \label{figConductanceMatrixSmoothSystemLongWire5TwoQuasiMBS}
\end{figure}

The final setup we consider is a topological nanowire with normal sections on both ends, see Fig.~\ref{figNanoWire1}, with parameter profiles specified in Fig.~\ref{figParameterProfileSmoothTwoQDot}.   
Such a system can host zero-energy quasi-MBSs at both ends of the nanowire in the topologically trivial regime. In App.~\ref{App:TopoNanowireAdditionalMaterial2}, we discuss the energy spectrum and the wavefunctions of bulk states, the latter is important for the understanding of the non-local conductances. The conductance patterns of the nanowire, depicted in Fig.~\ref{figConductanceMatrixSmoothSystemLongWire5TwoQuasiMBS}, exhibit features coming from the  left and right localized quasi-MBSs. As found previously, their conductance value is  quantized close to  $2e^2/h$.     The bulk-gap closing and reopening is  only weakly pronounced in the non-local  conductance $G_{RL}$, although a logarithmic color scale can reveal this process, see Fig.~\ref{figShortSmoothGapConductanceTemp40mKRightLeft5}. We note that even on the logarithmic scale the bulk states are poorly visible compared to Fig.~\ref{figShortSmoothGapConductanceTemp50mKRightLeft4}, whereas the energy spectrum (dark green dashed lines) clearly shows the bulk-gap closing and reopening.  This reduction of the non-local  conductance  signature of the bulk-gap closing and reopening by normal sections has been noted but not explained in Ref. \cite{Pan2021Three}. The reason for this reduction is that the bulk states have no support in the normal sections and especially the low energy states are confined to the middle of the superconducting section and therefore these bulk states have only very weak features in the local and non-local  conductances. Other states are extended throughout the whole nanowire and thus  contribute more strongly to the non-local  conductances, see App.~\ref{App:TopoNanowireAdditionalMaterial}. 

This suppression of the visibility of the bulk-gap closing in the non-local  conductance can be somewhat offset by decreasing the step height of the chemical potential at the interface between normal and superconducting section. Nonetheless, three-terminal experiments will require a very high resolution to measure the bulk-gap closing and reopening in superconducting nanowires with normal sections on both ends. 
If this gap behavior cannot be resolved experimentally, then it will also not be possible to distinguish MBSs from quasi-MBSs, even in long nanowires.  

\section{Conclusions \label{Sec:Conclusion}}

We analyzed transport properties of non-topological Rashba nanowires with normal sections that host ABSs. When the parameters of a normal section are close to a resonance condition and the ratio between the length of the superconductor and the ABS localization length is small, an ABS is pinned to zero energy over a wide range of Zeeman energies and has a finite probability density on both ends of the nanowire. The same effect occurs for the case of smooth spatial variation of  system parameters such as chemical potential and superconducting gap. As such, even though their origin is topologically trivial, calculations of local and non-local conductances reveal correlated ZBPs on the left and the right ends of the nanowire due to the ABSs. We conclude therefore that the measurement of correlated ZBPs on both ends of a superconducting nanowire is not an unambiguous  indicator for the presence of MBSs.

The observation  of the closing and reopening of the bulk-gap in the local and non-local conductances that should accompany a topological phase transition has also been considered in previous works as an additional indicator for the topological phase. However, we find here  that a second ABS at the other end of the nanowire can mimic the edge of the bulk-gap, when the ratio between the length of superconducting section and the localization length of the ABS is small. Therefore, local and non-local conductance measurements of ZBPs on each end with an apparent closing and reopening of the bulk-gap is also not an unambiguous  indicator for the presence of MBSs.

We conclude that, while next generation three-terminal experimental devices will have access to additional auxiliary features that can help clarify the origins of ZBPs, trivial ABSs can also generate conductance features similar to those expected from MBSs when such devices do not have long superconducting sections. In particular, we find that ABSs can produce correlated ZBPs and a feature reminiscent of a bulk-gap closing and reopening in local and non-local conductances. Our results therefore suggest that it is essential to perform measurements in systems with long superconducting sections and over a large region of parameter space if one wishes to gain confidence in a purported MBS signature. That said, ballistic transport experiments favor short nanowires since presently the production of devices with long mean free paths is challenging. It is therefore questionable whether current state-of-the-art or near-term Rashba nanowire devices will be able to conclusively rule out the effects of extended ABSs. Alternatively, these three-terminal detection methods should be supplemented by additional signatures observable in the bulk \cite{Gulden2016Universal,
Serina2018Boundary,Yang2019Robust,Tamura2019Odd,Sticlet2020All,Mashkoori2020Identification}
and related to the topological phase transition, such as the inversion of spin polarization in the lowest energy bulk states \cite{Szumniak2017Spin,Chevallier2018Topological}.

\section{Acknowledgements}
This project has received funding from the European Union's Horizon 2020 research and innovation programme under Grant Agreement No 862046 and under  Grant Agreement No 757725 (the ERC Starting Grant). This work was supported by the Georg H. Endress Foundation and the Swiss National Science Foundation.

\appendix
 \begin{widetext}
\section{Energy and Transport calculation} \label{App:DiffCond}
To obtain the energy spectrum and wavefunctions, we diagonalize numerically the Hamiltonian $H$. In addition to this we compute the 
differential conductance  $G_{\alpha\beta}=dI_{\alpha}/dV_{\beta}$ of the three-terminal device consisting of a nanowire with a grounded superconducting section and two normal leads at the left and right end using the Python package Kwant \cite{Groth2014Kwant}, which is based on the Blonder Tinkham Klapwijk (BTK) formalism \cite{Blonder1982Transition}.  In particular, we use Kwant to numerically calculate the $\rm S$-matrix and extract the transmission and reflection coefficients that determine the Andreev  conductance matrix  at zero temperature
\begin{align}
G_0=&\begin{pmatrix}
G_{LL,0}&G_{LR,0} \\
G_{RL,0}& G_{RR,0}
\end{pmatrix}=\frac{e^2}{h}\begin{pmatrix}
N_L-R_L^e(-eV_L)+A^e_L(-eV_L)&-T^e_{LR}(-eV_R)+A^e_{LR}(-eV_R) \\
-T^e_{RL}(-eV_L)+A^e_{RL}(-eV_L) & N_R-R_R^e(-eV_R)+A^e_R(-eV_R)
\end{pmatrix},
\end{align}
where $N_L$  ($ N_R $)  and $V_{L[R]}$  denote the number of channels and the gate voltage on the left (right) lead, respectively, $R_{\alpha}$ and $A_{\alpha}$ are the probabilities of an electron in lead $\alpha$ to be reflected as an electron or hole, respectively, and similarly, the coefficients $T_{\alpha\beta}$ and $A_{\alpha\beta}$ are the probabilities of an electron from lead $\beta$ to transmit as an electron or hole to lead $\alpha$, respectively.
The differential-conductance matrix  elements  \cite{DanonNonlocal2020,FregosoElectrical2013,LobosTunneling2014} at finite temperature $T$ are given by 
\begin{subequations}
\begin{align}
G_{LL}&=-\frac{e^2}{h}\int_{-\infty}^{\infty}\dd \omega\dv{f_L(\omega)}{\omega}\left[N_L-R_L^e(\omega)+A^e_L(\omega)\right], \\
G_{LR}&=\frac{e^2}{h}\int_{-\infty}^{\infty}\dd \omega\dv{f_R(\omega)}{\omega}\left[T^e_{LR}(\omega)-A^e_{LR}(\omega)\right], \\
G_{RL}&=\frac{e^2}{h}\int_{-\infty}^{\infty}\dd \omega\dv{f_L(\omega)}{\omega}\left[T^e_{RL}(\omega)-A^e_{RL}(\omega) \right], \\
G_{RR}&=-\frac{e^2}{h}\int_{-\infty}^{\infty}\dd \omega\dv{f_R(\omega)}{\omega}\left[N_R-R_R^e(\omega)+A^e_R(\omega)\right],
\end{align}
\end{subequations}
 where $ f_{L[R]} (\omega)= f (\omega +eV_{L[R]})$ denotes the Fermi distribution function 
 $f(\omega)=[\exp[\omega/(k_BT)]+1]^{-1}$,
with $ k_B $ being the Boltzmann constant. The temperature $T$
broadens peaks in the differential conductance. In this work, we perform the calculations using the temperature $ T=40 $ mK throughout, unless stated otherwise.
\end{widetext}

 \begin{figure}[!b]
\subfloat{\label{figShortNSNJunctionMajoranLeftConductance}\stackinset{l}{-0.02in}{t}{0.02in}{(a)}{\stackinset{l}{1.7in}{t}{0.02in}{(b)}{\stackinset{l}{-0.02in}{t}{1.17in}{(c)}{\stackinset{l}{1.7in}{t}{1.17in}{(d)}{\includegraphics[width=1\columnwidth]{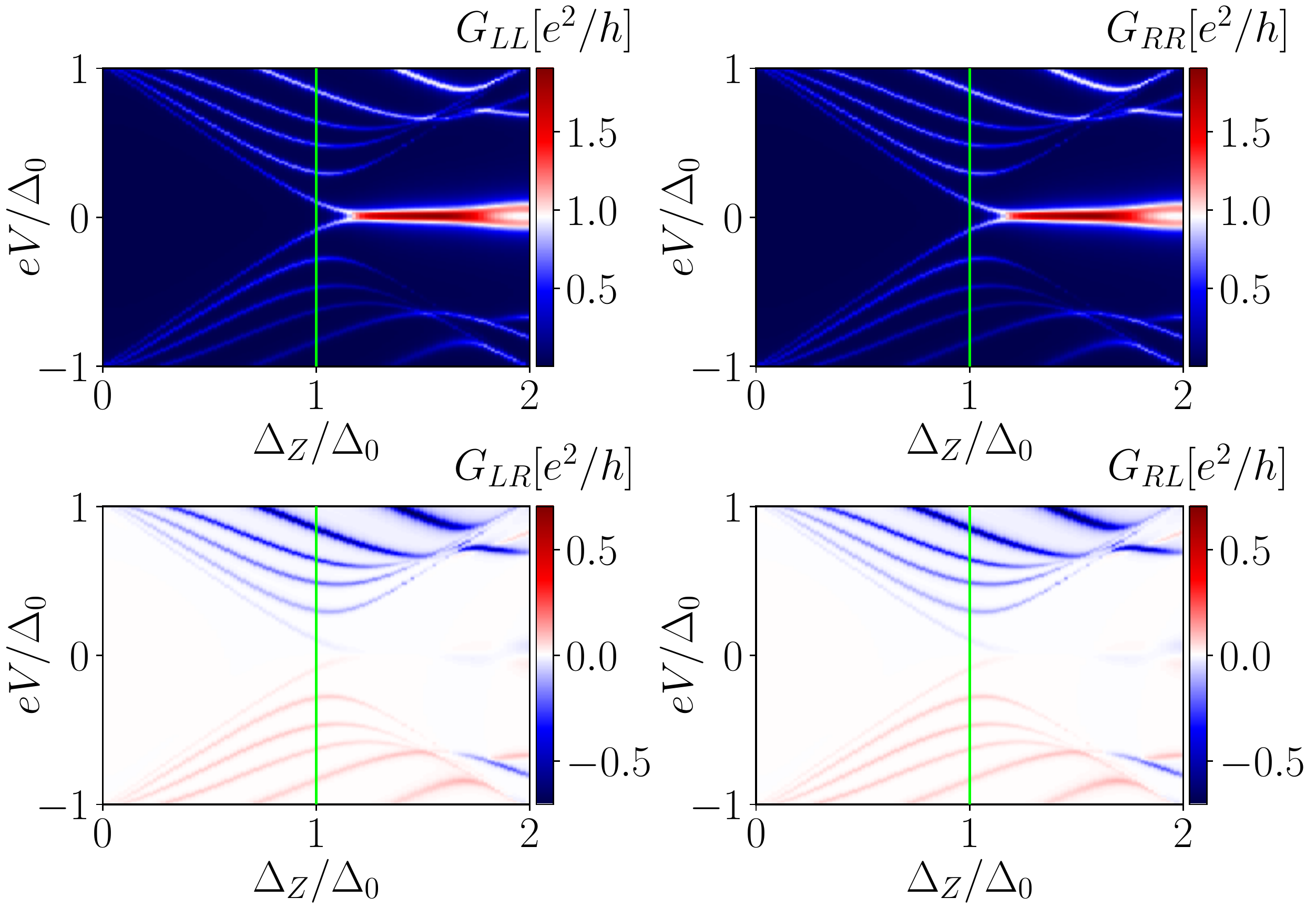}}}}}
}
\subfloat{\label{figShortNSNJunctionMajoranRightConductance}}
\subfloat{\label{figShortNSNJunctionMajoranLeftRightConductance}}
\subfloat{\label{figShortNSNJunctionMajoranRightLeftConductance}}
  \caption{Differential-conductance patterns in a short topological nanowire with uniform parameter profiles.   The local  conductances (a) $G_{LL} $   and (b) $G_{RR} $  are identical and exhibit ZBPs coming from the MBS in the topological phase. The height of this ZBP is a bit smaller than $2e^2/h$ due to thermal broadening.   The non-local  conductance (c) $G_{LR} $  and (d) $G_{RL}$   exhibit the  non-local bulk-gap closing and reopening process, close to the Zeeman energy $\Delta_Z=\Delta_0$, indicated by the green line.  Furthermore, the non-local  conductance exhibits features around zero energy originating from  overlapping MBSs.   The parameters are listed in Table~\ref{Tab:ParTopo} in App.~\ref{App:Parameters}. } 
 \label{figConductanceMatrixUniformShortWire1}
\end{figure}

 \section{Short uniform topological nanowire  \label{App:TopoNanowireAdditionalMaterial}}

In this section, we compute the  differential conductance of a short uniform nanowire, which enters a topological phase at $\Delta_Z=\Delta_0$, see Fig.~\ref{figConductanceMatrixUniformShortWire1}. This conductance behaviour is well known and is presented here in order to compare with that of a short non-topological nanowire which can exhibit similar signatures, see Fig.~\ref{figConductanceMatrixReegSystemTwoABSMimicTPT3}. The left  and right conductance patterns exhibit features coming from the MBSs after the topological phase transition. The MBSs overlap since their localization length is comparable to the system length and therefore the non-local  conductance also contains a weak MBS signature in this regime, see Figs.~\ref{figShortNSNJunctionMajoranLeftRightConductance} and \ref{figShortNSNJunctionMajoranRightLeftConductance}. A logarithmic scale can, however,  reveal these weak MBS signatures in the non-local  conductance, for example see also Fig.~\ref{figShortSmoothGapConductanceTemp40mKRightLeft5}.  

In short nanowires only a few states contribute to conductance at low biases close to the bulk-gap closing and reopening point. For instance, in the example shown in Fig.~\ref{figConductanceMatrixUniformShortWire1} only three states contribute. This should be compared to the conductance of the non-topological nanowire shown in Fig.~\ref{figConductanceMatrixReegSystemTwoABSMimicTPT3}, which hosts one state that mimics the bulk states undergoing a topological phase transition and is very similar to the behavior found in topological nanowires. In longer nanowires the energy level spacing between the bulk states decreases. As such, many states contribute to the conductance close to the bulk-gap closing and reopening point and therefore it is easier to distinguish between the bulk  and bound states. 

We note that the ZBP of the MBSs in the short topological nanowire is not quantized, which is also the case for the ZBP in the trivial nanowire. Experimentally, the robust quantization has not been observed so far. All in all,  a distinction between topological and trivial states in short nanowires via a local and non-local  conductance measurement is therefore challenging.

\section{Broadening of ZBP } \label{App:Broadening}
\begin{figure}[t]
\subfloat{\label{figBroadeningEffectTopoNW}\stackinset{l}{-0.02in}{t}{0.02in}{(a)}{\stackinset{l}{1.75in}{t}{0.02in}{(b)}{\includegraphics[width=1\columnwidth]{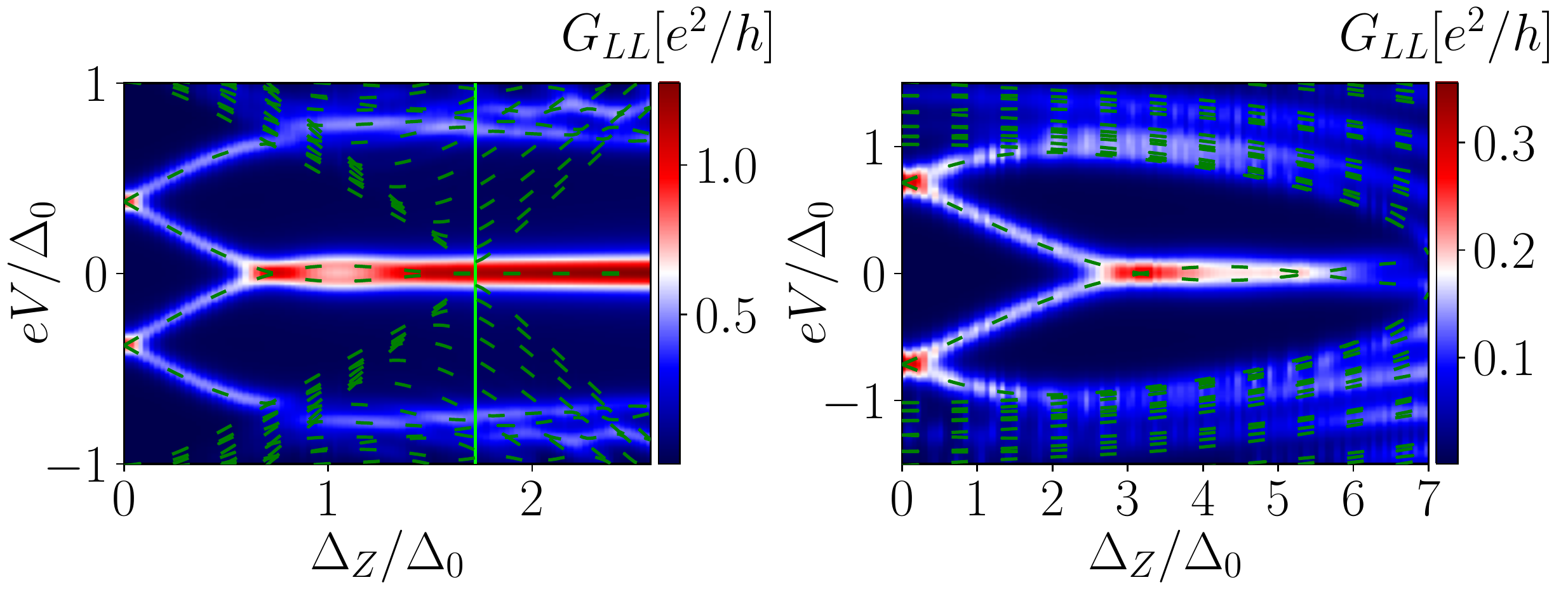}}}
}
\subfloat{\label{figBroadeningEffectNonTopoNW}}
  \caption{Strong broadening of the ZBP in a long  (a) topological  and (b) non-topological nanowire hosting an ABS. The differential conductance  peaks are in general broadened by different mechanisms such as strong coupling between leads and nanowire, external perturbations  due to the environment,  and high temperatures, here we effectively take these broadening mechanisms into account via a large effective temperature of $T=150$ mK. (a) The left local  conductance exhibits only a single ZBP, in contrast, the energies of the ABS (shown in dashed green lines) are not well pinned to zero. The strong broadening merges the two finite-energy peaks together to a single ZBP.  The topological phase transition is indicated by the green vertical line. (b)~The same effect is present in the non-topological system. The parameters are listed in Table~\ref{Tab:ParTopo} in App.~\ref{App:Parameters}.}
 \label{figBroadeningEffectForABS}
\end{figure}

We note that the calculated conductance peaks are relatively sharp.  In contrast,  experiments usually show broadened conductance patterns. Different mechanisms such as the strong coupling between leads and nanowire, external perturbations  due to environment effects, and high temperatures lead to a broadening of the conductance peaks.
In this section we consider long topological and non-topological nanowires, hosting  nearly zero-energy ABSs in the left normal section, and calculate the local  conductance $G_{LL}$, see Fig.~\ref{figBroadeningEffectForABS}. All broadening mechanisms are taken into account effectively via thermal effects, i.e. by choosing a relatively high temperature of $T=150$ mK.  
The resulting conductance is less sharp and is therefore in better agreement with the broader conductance features found in experiments. Furthermore, broadening prevents a high resolution mapping of the energy spectrum.  The left local  conductance $G_{LL}$ cannot resolve the fact that the ABS has a finite energy (in Fig.~\ref{figBroadeningEffectForABS} energies are shown as green dashed lines for comparison). The conductance peaks of the finite-energy ABS and its particle-hole partner are merged together into a single conductance peak at zero energy. As such, even in systems where  ABSs are not perfectly tuned to zero energy,  for example, if the resonance condition is fulfilled only approximately, an apparent ZBP in the conductance can still emerge.

  \section{ Absence of the signature of the bulk-gap closing in conductance}
  \label{App:TopoNanowireAdditionalMaterial2}
  \label{App:WavefunctionNoGapClosing}

   \begin{figure}[t]
\subfloat{\label{figQuasiMajoranaBothEndsLongWireEnergies}\stackinset{l}{0.0in}{t}{0.in}{(a)}{\stackinset{l}{1.73in}{t}{0.0in}{(b)}{\stackinset{l}{0.0in}{t}{1.2in}{(c)}{\stackinset{l}{1.73in}{t}{1.2in}{(d)}{\includegraphics[width=1\columnwidth]{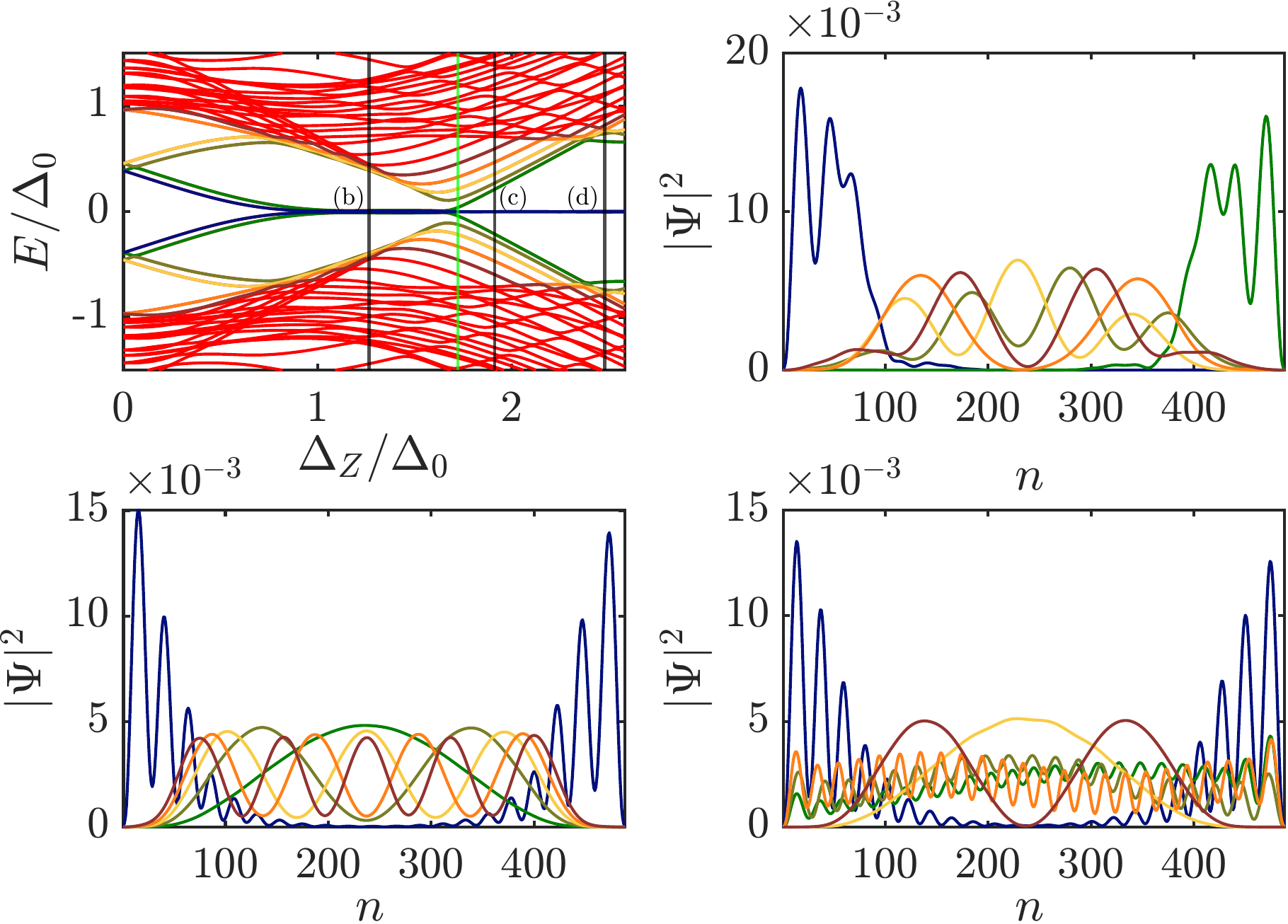}}}}}
}
\subfloat{\label{figQuasiMajoranaBothEndsLongWireProbDens1}}
\subfloat{\label{figQuasiMajoranaBothEndsLongWireProbDens2}}
\subfloat{\label{figQuasiMajoranaBothEndsLongWireProbDens3}}
 \caption{Topological nanowire hosting  quasi-MBSs on both ends.  (a) Energy spectrum and (b,c,d) probability densities of the quasi-MBSs and of the lowest bulk states at $\Delta_Z=\lbrace 1.26, 1.91,  2.48\rbrace\Delta_0$ [indicated by the black lines in  panel (a)]. The topological phase transition takes place at $\Delta_Z=1.73\Delta_0$ [indicated by the green line in  panel (a)]. The color of the energy states in panel (a) determines the color-code for their probability densities in the remaining panels.
 (b) In the trivial phase, the quasi-MBSs are well localized at the left and right end. The first four bulk states (khaki green, yellow, orange, dark red) are  mainly localized in the superconducting section of the nanowire. (c) Shortly after the topological phase transition, the wavefunctions of the energetically lowest  bulk states are still mainly localized within the superconducting section. (d) Deeply in the topological phase transition, the lowest bulk states (dark green and khaki green) are extended over the entire nanowire. These extended states are more visible in the non-local  conductance 
 (see Fig.~\ref{figConductanceMatrixSmoothSystemLongWire5TwoQuasiMBS}) than the bulk states shown in panel (b) and (c).
   The parameters are listed in Table~\ref{Tab:ParTopo} in App.~\ref{App:Parameters}. }
 \label{figQuasiMajoranaBothEndsLongWireProbDensities}
 \end{figure}
 
In this section we consider topological nanowires which host quasi-MBSs at both ends and analyze
the suppression of signatures of the topological phase transition in the conductance.
Our discussion focusses on long nanowires, for which elements of the corresponding conductance matrix is shown in Fig.~\ref{figConductanceMatrixSmoothSystemLongWire5TwoQuasiMBS}. The non-local conductances, $G_{RL}$ and $G_{LR}$, show only weak bulk-gap features at lower biases, despite the fact that the energy spectrum exhibits a clear bulk-gap closing and reopening consistent with  the topological phase transition, see Fig.~\ref{figQuasiMajoranaBothEndsLongWireProbDensities}. The phase transition is indicated by the green vertical line. This puzzle can be resolved by looking at the bulk wavefunctions,  
see Fig.~\ref{figQuasiMajoranaBothEndsLongWireProbDensities}.
The non-uniform chemical potential is responsible for confining the lowest energy bulk sub-gap states within the superconducting section. When the bulk gap closes in the superconducting section, the normal sections still nominally have a gap for states with nearly zero momentum originating from the interior branches of the spectrum.
In the trivial phase (see Fig.~\ref{figQuasiMajoranaBothEndsLongWireProbDens1}), the quasi-MBSs  (blue, dark green)  are well localized at the left and right ends of the nanowire. As a result, they couple strongly to the leads.
In contrast, the energetically lowest bulk states (khaki green, yellow, orange and dark red) are mainly localized within the superconducting section. 
 Thus, there is hardly any coupling to the leads and, as such, these bulk states only weakly contribute to the non-local  conductance of the trivial phase.

Right after the topological phase transition (see Fig.~\ref{figQuasiMajoranaBothEndsLongWireProbDens2}), the wavefunctions of the energetically lowest  bulk states are also mainly localized within the superconducting section and not in the normal sections.  This results in a similar absence of a corresponding non-local  conductance signal  as occurred in the trivial phase. In general, we find the lower the energy of the bulk state the more it is localized within the superconducting section.  For example, the energetically lowest state (dark green) is more localized than the fifth bulk state (dark red). Furthermore, these bulk states are spatially  separated from the left and right  ends of the nanowire, so that a local  conductance measurement also can not resolve such states. In contrast, the MBSs (dark blue) are mainly localized in the  normal sections and decay into the superconductor making them highly visible in local  conductance measurements.

 \begin{figure}[t]
\subfloat{\label{figLongSmoothGapConductanceTemp40mKLeft6}\stackinset{l}{-0.00in}{t}{-0.0in}{(a)}{\stackinset{l}{1.67in}{t}{-0.0in}{(b)}{\stackinset{l}{-0.0in}{t}{1.08in}{(c)}{\stackinset{l}{1.67in}{t}{1.08in}{(d)}{\stackinset{l}{-0.0in}{t}{2.16in}{(e)}{\stackinset{l}{1.67in}{t}{2.16in}{(f)}{\includegraphics[width=1\columnwidth]{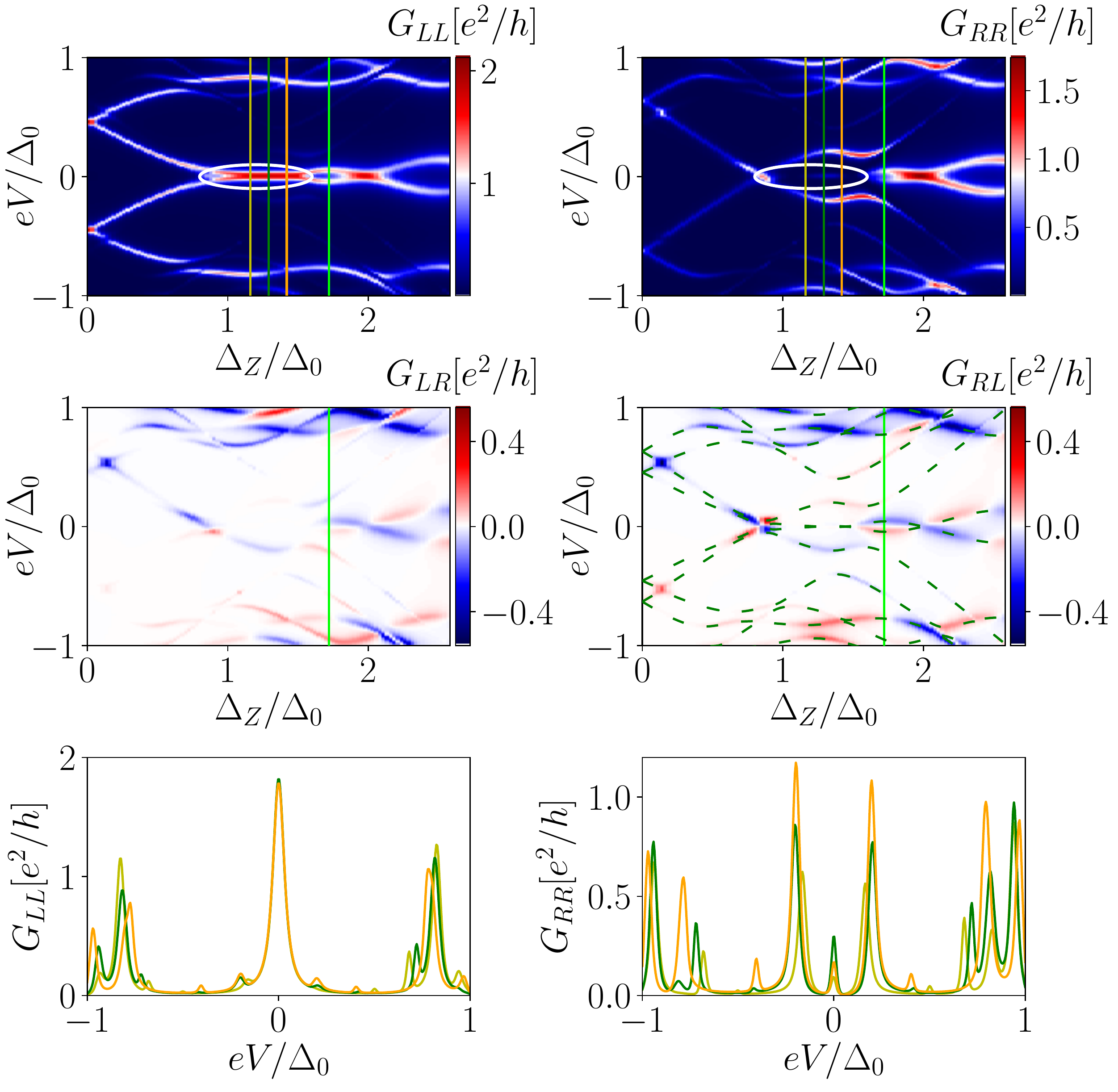}}}}}}}
}
\subfloat{\label{figLongSmoothGapConductanceTemp40mKRight6}}
\subfloat{\label{figLongSmoothGapConductanceTemp40mKLeftRight6}}
\subfloat{\label{figShortSmoothGapConductanceTemp40mKRightLeft6}}
\subfloat{\label{figShortWireLineCutGLL3}}
\subfloat{\label{figShortWireLineCutGRR3}}
  \caption{Same as Fig.~\ref{figShortWireSmoothParameterConductance} but for the nanowire containing an additional right normal section hosting an ABS, which mimics a bulk-gap closing and reopening in the non-local  conductances close to $\Delta_Z\approx0.9\Delta_0$.  The left quasi-MBSs are visible in the  local  conductances (a)  $G_{LL} $   and (b) $G_{RR} $.  However, in $G_{RR}$ they are less pronounced. The non-local  conductances (c) $G_{LR} $  and (d)  $G_{RL}$ contain features coming from the lowest bulk states as well as from the right ABS, which leaks through the superconducting section.  Line cuts (e,f) of the local  conductances  $G_{LL} $ and $G_{RR} $ at the Zeeman energies $\Delta_Z=\lbrace1.16, 1.29, 1.42\rbrace\Delta_0$ [indicated by the yellow, dark green and orange lines in (a,b)] both contain features coming from the left quasi-MBSs. The parameters are listed in Table~\ref{Tab:ParTopo} in App.~\ref{App:Parameters}. }
 \label{figConductanceMatrixSmoothSystemShortWireTwoQuasiMBS}
\end{figure}

Deep inside the topological phase (see Fig.~\ref{figQuasiMajoranaBothEndsLongWireProbDens3}), the lowest bulk states (dark green and khaki green) originating  from the exterior branches of the spectrum (from finite Fermi momentum) are extended over the entire nanowire. These delocalized states couple strongly to the leads and do contribute to the non-local  conductance. In contrast, some of the energetically higher states (such as the yellow and dark red) originating  from the interior branches of the spectrum (from nearly zero Fermi momentum), -- which are related to the states discussed in Fig.~\ref{figQuasiMajoranaBothEndsLongWireProbDens2} -- remain confined in the superconducting section and therefore contribute less to the non-local  conductance. 

The absence of a clear bulk-gap closing and reopening signal in such a setup makes it essentially impossible to determine the location of the topological phase transition measuring local and non-local  conductances and therefore it is also not possible to conclusively determine whether the system hosts MBSs or two quasi-MBSs. Although discussed here for long topological nanowires, this behavior also occurs in short topological nanowires.

\section{Interplay between quasi-MBSs at the left end and ABS at the right end of a short topological nanowire} \label{App:QuasiMBSAndABS}

Finally, we consider a short topological nanowire with quasi-MBSs on the left end and an ABS on the right end. The ABS is again tuned so that it mimics a bulk-gap undergoing  a topological phase transition.  The parameter profiles of superconducting gap and chemical potential are not identical in two normal sections. We choose a smooth parameter profile at the interface between $\text{N}_1$ and S, and a step-like profile at the interface between   S and $\text{N}_2$. 
 The elements of the conductance matrix are shown in Fig.~\ref{figConductanceMatrixSmoothSystemShortWireTwoQuasiMBS}.  The energy spectrum (dark green dashed lines) agrees  well with features in the non-local  conductance $G_{RL}$. The left quasi-MBSs leak through the superconducting section and generates a small ZBP in the right conductance $G_{RR}$, see Fig.~\ref{figLongSmoothGapConductanceTemp40mKRight6}. This behavior is similar to the one of the setup shown in Fig.~\ref{figShortWireSmoothParameterConductance} and is again explained  by the extended nature of wavefunctions. The ZBP originating from the left quasi-MBSs in the right local  conductance is more pronounced in linecuts, see Fig.~\ref{figShortWireLineCutGRR3}.  
The energy of the right ABS decreases with increasing Zeeman energy until it is nearly zero at the same values of the magnetic field at which the quasi-MBSs  begin to be pinned to zero energy ($\Delta_Z\approx 0.9\Delta_0$). At stronger magnetic fields ($0.9\Delta_0<\Delta_Z< 1.4\Delta_0$), the right ABS moves away from zero energy,  mimicking the reopening of the bulk gap. The true topological phase transition, however, takes place only around $\Delta_Z\approx 1.74\Delta_0$.  The right ABS is not only visible in $G_{RR} $ but also in the non-local  conductances, see Fig.~\ref{figLongSmoothGapConductanceTemp40mKLeftRight6} and \ref{figShortSmoothGapConductanceTemp40mKRightLeft6}. Additionally this ABS generates  a small feature in $G_{LL} $ which is only visible in the linecut shown in Fig.~\ref{figShortWireLineCutGLL3}. We note that the height of this right ABS peak in $G_{LL} $   is comparable with the one of the energetically lowest  bulk state. We conclude that, in experiments, an ABS on the right end could easily mask a topological phase transition. 

\begin{widetext}
\section{Parameter values} \label{App:Parameters}
In this section, we list all parameters used in each figure, see Table \ref{Tab:ParNonTopo1} and Table \ref{Tab:ParTopo23}. The hyphen in the table indicates that the respective parameter was not included in the calculation: For example the nanowire considered in Fig.~\ref{figABSSingleQdotEnergySpectrumMagDependence1HugeGap}  does not include a second normal section to the right of the superconducting section. Furthermore, the asterisks $\ast$ indicates that the corresponding parameter runs over a finite interval which is indicated in the figure. 
The parameters from Figs.~\ref{figSingleABSResoannceProbDensMagFieldDependencelongWirePinningRegion} and  \ref{figSingleABSResoannceProbDensMagFieldDependenceshortWirePinningRegion}  are the same as the ones from Figs.~\ref{figABSSingleQdotEnergySpectrumMagDependence1HugeGap} and \ref{figABSSingleQdotEnergySpectrumMagDependence1ShortWire}.
We choose a temperature of $T=40$ mK in all plots except in Fig.~\ref{figBroadeningEffectForABS}, where we take $T=150$ mK. Furthermore, the effective lattice constant is  $a=5$ nm in all plots. All energy values in the following tables are given in units of meV.

\begin{table}[t]
 \centering
\caption{ Parameters used to model non-topological nanowires\label{Tab:ParNonTopo1}}
 \centering
\begin{tabular}{c|ccccccccccccccccccccccccccccc}
\hline
\hline
Fig.&$N_1$ &$N_2$ &$N_S$ &$N_{B,1}$ &$N_{B,2}$ &$t_1$ &$t_2$ &$t_S$&$\mu_1$ &$\mu_2$ &$\mu_S$ &$\Delta_0$ &$\Delta_{Z}$&$\Delta_{Z}^c$&$\alpha_1$ &$E_{so,1}$ &$\alpha_2$  & $E_{so,2}$ &$\gamma_1$ &$\gamma_2$ &$\mu_{L}$ &$\mu_{R}$ \\

 \hline
\ref{figEnergyOscillationResonance09DeltaZc} & 60&-&400&-&-&100&-&20&$\ast$&-&2&0.25&1.58&1.75&$\ast$&$\ast$&-&-&-&-&-&-\\
 \ref{figEnergyOscillationResonance075DeltaZc} & 60&-&400&-&-&100&-&20&$\ast$&-&2&0.25&1.31&1.75&$\ast$&$\ast$&-&-&-&-&-&-\\
\ref{figFiniteChemicalPotentialZeroBiasPinning}& 60&-&400&-&-&100&-&20&0.44&-&2&0.25&$\ast$&1.75&13.35&1.78&-&-&-&-&-&-\\
\ref{figFiniteChemicalPotentialZeroBiasPinningProblemPhaseShift}& 60&-&400&-&-&100&-&20&0.3&-&2&0.25&$\ast$&1.75&7.97&0.63&-&-&-&-&-&-\\
\ref{figABSSingleQdotEnergySpectrumMagDependence1HugeGap}& 60&-&400&-&-&100&-&20&0&-&2&0.25&$\ast$&1.75&14.35&2.06&-&-&-&-&-&-\\
\ref{figABSSingleQdotEnergySpectrumMagDependence1}& 60&-&400&-&-&100&-&20&0&-&2&0.09&$\ast$&1.75&14.35&2.06&-&-&-&-&-&-\\
\ref{figABSSingleQdotEnergySpectrumMagDependence1ShortWire}& 60&-&175&-&-&100&-&20&0&-&2&0.25&$\ast$&1.75&14.35&2.06&-&-&-&-&-&-\\
 \ref{figConductanceMatrixReegSystem1} & 90&7&140&7&7&100&100&20&0&0&2&0.25&$\ast$&1.75&13.75&1.89&13.75&1.89&10&10&20&20\\
\ref{figABSTwoQdotEnergySpectrumMagDependenceBothInResonance} & 60&60&400&-&-&100&100&20&0&0 &2&0.25&$\ast$&1.75&14.35&2.06&14.35&2.06&-&-&-&-\\
\ref{figABSTwoQdotEnergySpectrumMagDependenceOneInResonance}& 60&70&400&-&-&100&100&20&0&0&2&0.25&$\ast$&1.75&14.35&2.06&14.35&2.06&-&-&-&-\\
\ref{figTwoQDotsOneABSMimicsGapclosingLongtWire2} & 60&40&400&-&-&100&100&20&0&0&2&0.25&$\ast$&1.75&14.35&2.06&10.33&1.07&-&-&-&-\\
\ref{figTwoQDotsOneABSMimicsGapclosingshortWire4}& 90&30&140&7&7&100&100&20&0&0&2&0.25&$\ast$&1.75&13.75&1.89&2.75&0.08&10&10&-&-\\
\ref{figConductanceMatrixReegSystemTwoABSMimicTPT3}& 90&30&140&7&7&100&100&20&0&0&2&0.25&$\ast$&1.75&13.75&1.89&2.75&0.08&10&10&20&20\\
\ref{figBroadeningEffectNonTopoNW} & 60&7&400&7&7&100&100&20&0.3&0&2&0.25&$\ast$&1.75&8.80&0.77&8.80&0.77&10&5&5&5\\
\hline
\hline
\end{tabular}
\end{table}

\begin{table}[b]
\caption{ Parameters used to model topological nanowires \label{Tab:ParTopo23} \label{Tab:ParTopo} }
 \centering
\begin{tabular}{c|cccccccccccccccccccccc}
\hline
\hline
Fig. & $N_1$ & $N_2$ & $N_S$ &   $N_{B,1}$ & $N_{B,2}$&$t$ &$\mu_1$ &$\mu_2$ &  $\mu_S$ & $\Delta_0$ & $\Delta_{Z}^c$ & $\alpha$ & $E_{so}$ & $\gamma_1$ & $\gamma_2$& $\mu_{L}$& $\mu_{R}$& $\lambda_{S,L}$& $\lambda_{S,R}$& $\lambda_{L}$& $\lambda_{R}$\\
\hline
\ref{figLongWireSmoothGapandPotEnergies}  & 40 & 4 & 400 & 4 &4  &102 & 0.2 & 0.2& 0.7&0.5&- &3.5& 0.12& 10 &10 &- &-&20&-&20 &- \\
\ref{figShortWireSmoothGapandPotEnergies} & 40 & 4 & 120 & 4 &4  &102 & 0.2 & 0.2& 0.7&0.5&- &3.5& 0.12& 10 &10 &- &-&20&-&20 &- \\
\ref{figConductanceMatrixSmoothSystemShortWire4}& 40 & 4 & 400 & 4 &4  &102 & 0.2 & 0.2& 0.7&0.5&- &3.5& 0.12& 10 &10 &20 &20&20&-&20 &- \\
\ref{figShortWireSmoothParameterConductance} & 40 & 4 & 120 & 4 &4  &102 & 0.2 & 0.2& 0.7&0.5&- &3.5& 0.12& 10 &10 &20 &20&20&-&20 &- \\
\ref{figConductanceMatrixSmoothSystemLongWire5TwoQuasiMBS}  & 40 & 48 & 400 & 4 &4  &102 & 0.2 & 0.1& 0.7&0.5&- &3.5& 0.12& 10 &10 &20 &20&20&24&20 &24 \\
\ref{figConductanceMatrixUniformShortWire1} & 4 & 4 & 140 & 4 &4  &102 & 0 & 0& 0&0.7&- &3.5& 0.12& 10 &10 &20 &20&-&-&- &- \\
\ref{figBroadeningEffectTopoNW}  & 40 & 4 & 400 & 4 &4  &102 & 0.2 & 0.2& 0.7&0.5&- &3.5& 0.12& 10 &10 &20 &20&10&-&20 &- \\
\ref{figQuasiMajoranaBothEndsLongWireProbDensities} & 40 & 48 & 400 & 4 &4  &102 & 0.2 & 0.1& 0.7&0.5&- &3.5& 0.12& 10 &10 &- &-&20&24&20 &24 \\
\ref{figConductanceMatrixSmoothSystemShortWireTwoQuasiMBS} & 40 & 32 & 120 & 4 &4  &102 & 0.2 & 0.2& 0.7&0.5&- &3.5& 0.12& 10 &10 &20 &20&20&-&20 &-\\
\hline
\hline
\end{tabular}
\end{table}

\end{widetext}

\bibliographystyle{apsrev4-1}
\bibliography{Literatur2}
\end{document}